\documentclass[aps,pre,groupedaddress,longbibliography]{revtex4-2}
\usepackage{xcolor}
\usepackage{graphicx}
\usepackage{amssymb}
\usepackage{amsmath}
\usepackage[english]{babel}
\usepackage{epstopdf, epsfig}
\usepackage{hyperref}

\begin{document}

	\title{Dynamic behavior of elastic strips near shape transition}

	\author{Basile Radisson and Eva Kanso\footnote{Corresponding author:Kanso@usc.edu}}
	
	\affiliation{Department of Aerospace and Mechanical Engineering, \\ University of Southern California, Los Angeles, CA 90089-1191, USA}

	\date{\today}

	\begin{abstract}
       Elastic strips provide a canonical system for studying the mechanisms governing elastic shape transitions.  Buckling, linear snap-through, and nonlinear snap-through have been observed in boundary-actuated strips and linked to the type of bifurcation the strip undergoes at the transition. For nonlinear snap-through, previous work obtained the normal form at the bifurcation.   
        However, to date, there is no methodology for extending this analysis to other types of transition.
        Here, we study a set of three systems where a buckled elastic strip is actuated through rotation of its boundaries. Depending on the direction of rotation, the system exhibits all three types of shape transitions.
      We introduce a simple method to analyse the dynamic characteristics of an elastic structure near a transition. This method allows us to extend, in a straightforward manner, the asymptotic analysis proposed for nonlinear snap-through to the two other types of transition. We obtain the normal forms of these bifurcations, and show how they dictate all the dynamic characteristics of the elastic strip.
       This analysis provides a profound understanding of the physical mechanisms governing elastic shape transitions and reliable tools to diagnose and anticipate these transitions.
	\end{abstract}

	
	\maketitle	
	\section{Introduction}\label{sec:introduction}
	
	Elastic shape transitions arise when an elastic structure is in an equilibrium configuration that becomes unstable or suddenly disappears, under variation of a control parameter. These transitions are commonly classified into \textit{buckling} and \textit{snap-through}.  Buckling corresponds to a supercritical transition where, for an infinitesimal variation of the control parameter, the elastic structure moves by an infinitesimal amount \cite{nayfeh2008}. Snap-through corresponds to a subcritical transition where an infinitesimal variation of the control parameter induces a finite motion of the elastic structure  \cite{goriely2006, gomez2017, siefert2022}.
	In other words, buckling occurs when, at the bifurcation point, the system  transitions smoothly to a newly created stable equilibrium branch; snap-through occurs when the structure must jump to a distant equilibrium. The study of these transitions consists of determining (i) the value of the control parameter at which the bifurcation takes place, (ii) the number of branch of solutions that split off at the bifurcation point, and (iii) the behavior of these solutions in the neighborhood of the bifurcation point. 
	
	Once (i) is known, a common way to solve (ii) and (iii) at once is to reduce the dynamic of the elastic structure near the bifurcation to the temporal evolution of the amplitudes of \textit{critical normal modes}; these are modes whose eigenfrequency vanishes at the bifurcation and that are only mildly unstable or slightly damped around the bifurcation.  All other modes are strongly damped and rapidly attenuated and play only a marginal role in the dynamics {near the bifurcation}. The  \textit{amplitude equations} or \textit{normal forms} describe the behavior of these critical modes and give a good approximation of the dynamics near the transition \cite{goriely1996, goriely2000, gomez2017, gomez2018b, kodio2020}. Importantly, they allow precise classification of these transitions according to the type of bifurcation exhibited by the \textit{normal forms}.  
	
 Elastic strips have been used in recent years as a canonical system to obtain a fundamental understanding of elastic shape transitions \cite{pandey2014, gomez2017, gomez2018b, sano2017, sano2018, sano2019, librandi2020}. Specifically, \textit{buckling} is observed under transversal shearing of a clamped-clamped or hinged-hinged strip~\cite{sano2018}); \textit{linear snap-through}, which denotes snap-through transitions where the early dynamics is linear, is observed when the both ends of a clamped-clamped strip are rotated symmetrically~\cite{gomez2017}); \textit{nonlinear snap-through}, which denotes snap-through transition where the dynamics is nonlinear even at early time, is observed under asymmetric rotational actuation of the clamped-clamped strip~\cite{gomez2017} or transverse actuation of clamped-hinged strip~\cite{sano2018}. 
 
 For \textit{nonlinear snap-through}, Gomez et al. \cite{gomez2017, gomez2018, gomez2018b} obtained the normal form of the system near the transition and showed that it corresponds to a saddle-node bifurcation. This allowed them to explain all the dynamic properties of this transition. In particular, they related the abnormally slow dynamics of snap-through - which was commonly attributed to dissipation mechanisms and/or viscoelastic effects - to the critical slowing down near the bifurcation. The normal form was obtained using reduction order methods that rely on an asymptotic analysis in the vicinity of the bifurcation. This analysis requires to expand the different variables involved in the problem in terms of the bifurcation parameter, with correct scaling.  In \cite{gomez2017, gomez2018, gomez2018b}, these scalings were postulated and justified a posteriori.
Yet, to date, there is no  systematic way to obtain these scalings in different systems. 

In this paper,  we propose a systematic way to obtain the scalings of the different variables near the bifurcation. This allows us to expand the analysis of Gomez \textit{et al.} to the two other types of transition identified in the literature (\textit{buckling} and \textit{linear snap-through}).
 We study  three  systems inspired by \cite{gomez2017}.  A buckled elastic strip, clamped at both ends, is boundary actuated by rotating one or both of its boundaries. Depending on the direction around which we rotate the boundaries, we get three  systems that exhibit the three types of elastic shape transitions reported in the literature: \textit{buckling}, \textit{linear snap-through} and \textit{nonlinear snap-through}.
     
We analyze in details the static equilibria and the dynamics of these three systems numerically, by solving the fully nonlinear discrete Cosserat rod equations~\cite{cosserat1909, gazzola2018}, and analytically, in the context of the quasi-linear geometrically constrained Euler Beam model~\cite{pandey2014, gomez2017}. 
From this analysis, we develop a systematic approach to obtain the scaling of the different variables near the bifurcation directly from data. This allows us to extend the analysis carried out in \cite{gomez2017} and derive the normal form of the bifurcation for the two other types of shape transitions. We demonstrate that all the dynamic characteristics of the elastic structure are dictated by the nature of the underlying bifurcation. In particular, we show that the scaling of dynamic variables with the bifurcation parameter provides a robust marker of the type of shape transition the system undergoes. This analysis leads to reliable  tools for the diagnosis of shape transitions in elastic structures. We conclude by showing how these tools can be exploited to anticipate shape transitions.

\begin{figure}[t]
 	\centering
 	\includegraphics[width =\textwidth]{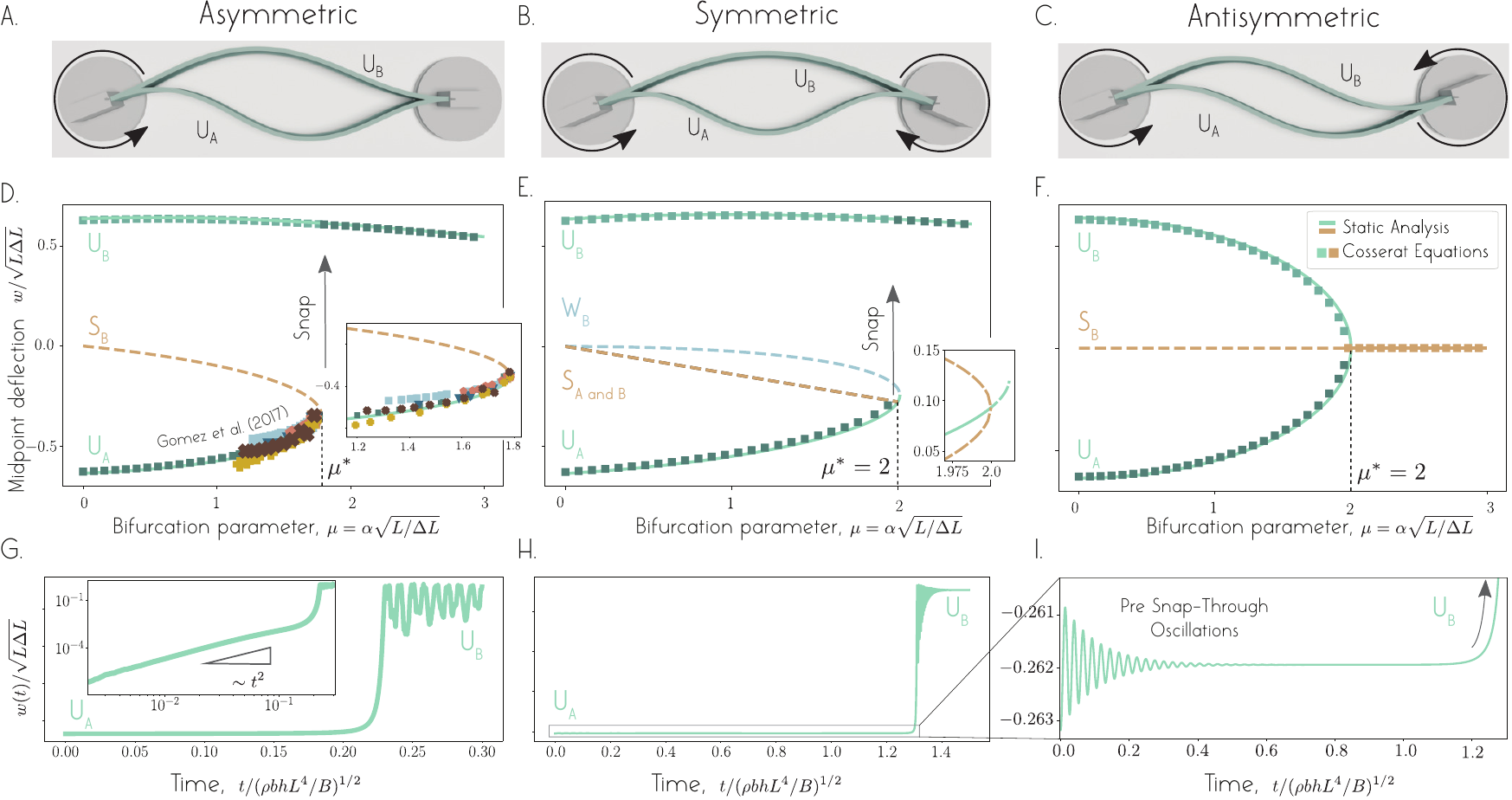}
 	\caption{\footnotesize{\textbf{Rotational boundary actuation} When the boundaries of a buckled strip are rotated by an angle $\mu$ in an (A) asymmetric , (B) symmetric , or (C) antisymmetric  fashion, the two buckling configurations U\textsubscript{A} and U\textsubscript{B} are modified and approach each other until they merge in a single configuration. (D, E, F) The transition from two stable equilibria to a single equilibrium depends on the boundary actuation. This is depicted by plotting the evolution of the mid-point of the strip in term of $\mu$ for each equilibrium. The transition is 
  abrupt for the asymmetric and symmetric cases (D and E) and smooth for the antisymmetric case (F). In D and E, the strip snaps from U\textsubscript{A} to U\textsubscript{B}. (G) The snap-through dynamic is monotonic for the asymmetric case  and (H, I) preceded by damped oscillations for the symmetric case; see Appendix~\ref{app:sym_osc} for an analysis of the origin of these oscillations.}}
 	\label{fig:fig2}
 \end{figure}

\section{Numerical observations based on the 3D Cosserat rod theory}
\label{sec:numerics}

We consider an elastic strip of length $L$ and rectangular cross-section of width $b$ and thickness $h$ that is clamped at both ends and strain-free in a straight reference configuration. The strip material properties are characterized by its density $\rho$ and Young's modulus $E$. The straight strip is first compressed longitudinally, by constraining its end-to-end distance $L-\Delta L$ to be shorter than the strip length $L$. This causes the strip to buckle following a supercritical pitchfork bifurcation, known as the Euler-buckling instability (e.g.,\cite{nayfeh2008, howell2009}). The buckled strip admits two equally-likely, symmetric buckled states. The bistable elastic structure is driven through shape transition by rotating either one or both ends by a non-zero angle $\alpha$ (Fig.~\ref{fig:fig2}). Rotational boundary actuation leads to another bifurcation, as $\alpha$ increases, where the system transitions from bistable to monostable. The nature of this bifurcation and the dynamic behavior of the strip around that bifurcation are the main topics of this study.

\begin{figure}[t]
 	\centering
 	\includegraphics[width =\textwidth]{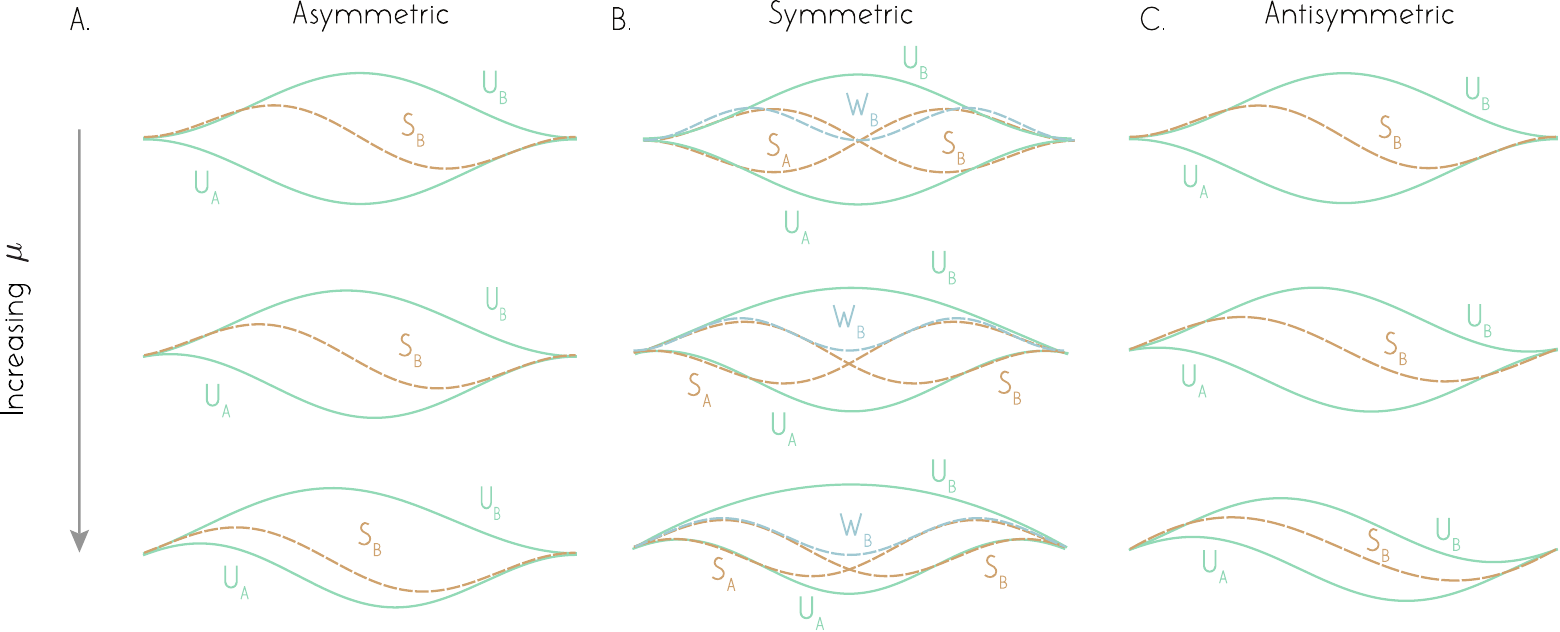}
 	\caption{\footnotesize{\textbf{Evolution of the equilibrium shapes.} Evolution of the equilibrium shapes obtained from the Euler beam model under (A) asymmetric, (B) symmetric and (C) antisymmetric boundary actuation when approaching the bifurcation. For the Asymmetric case, U\textsubscript{A} monotonically approaches S\textsubscript{B} until they merge and both disappear (Fig. \ref{fig:fig2}D). For the Symmetric case, U\textsubscript{A} approaches S\textsubscript{A} and S\textsubscript{B} until they all merge in an unstable equilibrium. Soon after the equilibrium born from the merging of these three shapes merges with with W\textsubscript{B} and both disappear (Fig. \ref{fig:fig2}E). For the Antisymmetric case U\textsubscript{A}, S\textsubscript{A}, and S\textsubscript{B} approach each other until they all merge in a single equilibrium (Fig. \ref{fig:fig2}F).}}
 	\label{fig:fig2Bis}
 \end{figure}

We numerically investigate the equilibrium configurations of the boundary-actuated strip using an implementation of the discrete Cosserat Rod theory \cite{gazzola2018}.  Starting from the clamped-clamped straight strip, we quasi-statically push the two ends towards each other, waiting sufficiently long after each decrement for the strip to reach mechanical equilibrium, until the strip buckles into one of two energetically identical states and a targeted end-to-end distance $L-\Delta L$ is reached. 
 The two stable Euler buckled states, hereafter denoted by U\textsubscript{A} and U\textsubscript{B}, correspond to the first buckling mode. 
 We then subject the buckled strip to rotation of one or both of its clamped boundaries. We consider asymmetric, symmetric, and  antisymmetric boundary rotations (Fig.~\ref{fig:fig2}). In the asymmetric case, we rotate one end while holding the other at zero angle (Fig.~\ref{fig:fig2}A) as done experimentally in~\cite{gomez2017}. In the symmetric case, both ends are rotated by an equal amount in opposite directions (Fig.~\ref{fig:fig2}B), while in the antisymmetric case, both ends are rotated by an equal amount in the same direction (Fig.~\ref{fig:fig2}C).  As we rotate the strip's endpoints,   U\textsubscript{A} and U\textsubscript{B} morph into two new stable equilibrium shapes (Fig.~\ref{fig:fig2Bis}).

To facilitate later analysis, we report the strip's behavior in non-dimensional form. Following \cite{gomez2017}, we scale time $t$ by the elastic time scale $ \sqrt{\rho b h L^4/B}$, longitudinal distance by the strip length $L$, and the strip transverse deflection  by $\sqrt{L\Delta L}$. We use $w(s,t)$ to refer to the strip's transverse deflection in the $y$-direction, consistent with the notation in the Euler-Bernoulli theory introduced in~\S\ref{sec:EB}. In non-dimensional form, the angle imposed at the boundaries becomes $\mu= \alpha \sqrt{{L}/{\Delta L}}$, which was first introduced in \cite{gomez2017}. The parameter $\mu$ balances the slope $\alpha$ imposed at the boundary with the natural slope $\sqrt{\Delta L/L}$ adopted by the strip in order to satisfy the end-to-end shortening. 

In Fig.~\ref{fig:fig2} and \ref{fig:fig2Bis}, we vary $\mu$ by holding $\Delta L$ fixed and varying $\alpha$. For all three types of boundary actuation, the two initially-stable buckled states U\textsubscript{A} and U\textsubscript{B} get modified until a threshold value $\mu^*$ is reached (Table~\ref{tab:muStarNumRot}). Above this threshold, only one stable equilibrium configuration is available for the strip. The way the system transitions from two equilibrium states to one depends on the type of boundary actuation.

\begin{table}
\caption{\textbf{Threshold values $\mu^\ast$ of the bifurcation parameter} $\mu$ obtained numerically using discrete Cosserat simulations for $\Delta L/L=10^{-2}$ and (semi)-analytically using the Euler beam model. Values without decimal are analytically exact while values with decimals are approximate.}
\begin{tabular}{c| ccc}
& \textbf{Asymmetric} & \textbf{Symmetric} & \textbf{Antisymmetric}\\[2mm]
\toprule
Numerical $\mu^\ast$&$1.763$&$1.973$&$1.967$\\[2mm]
Analytical $\mu^\ast$&$1.782$&$2$&$2$\\
\hline
\end{tabular} 
\label{tab:muStarNumRot}
\end{table}

In the asymmetric case, the transition happens at $\mu^*\approx1.763$. 
 The stable branch corresponding to the inverted shape U\textsubscript{A} disappears suddenly at $\mu^*$ and only the natural shape U\textsubscript{B} remains available for $\mu > \mu^*$.
 This is evidenced by plotting the evolution of the midpoint deflection $w$ for the two equilibrium states U\textsubscript{A} and U\textsubscript{B} as a function of the actuation parameter $\mu$ (Fig. \ref{fig:fig2}D square symbols).
 At the transition $\mu^\ast$, an infinitesimal variation $\delta \mu$ causes sections of the strip that is initially in the U\textsubscript{A} configuration to move by a finite amount before reaching the equilibrium in the U\textsubscript{B} configuration. This is typical of a \textit{snap-through} transition. Near the transition, the equilibrium configurations obtained in our numerics compare well with the experimental data obtained in~\cite{gomez2017} (inset Fig \ref{fig:fig2}D).
  The time-evolution of the snapping event is shown in Fig. \ref{fig:fig2}L by plotting  the midpoint deflection versus time right after passing the threshold $\mu^*$. In the inset of Fig. \ref{fig:fig2}G, we plot the evolution of the quantity $w(s=L/2, t)-w(s=L/2,t=0)$ on a logarithmic scale in order to observe how the strip goes away from its initial configuration  during the snapping event. Clearly, the strip monotonically moves from U\textsubscript{A} to U\textsubscript{B} in an algebraic manner, as observed and explained in \cite{gomez2017}.

    In the symmetric case,  the branch corresponding to the U\textsubscript{A} configuration suddenly disappears at $\mu^*\approx1.973$, and, thereafter only U\textsubscript{B} is observed (Fig. \ref{fig:fig2}E square symbols). When the strip in the inverted  U\textsubscript{A} configuration reaches the end of this branch, it has to snap to the other configuration.
Interestingly, when the strip in the U\textsubscript{A} configuration is pushed beyond the threshold $\mu^*$, the strip first exhibits damped oscillations and then suddenly snaps to the U\textsubscript{B} configuration as shown in Fig. \ref{fig:fig2}H and \ref{fig:fig2}I. This differs from the transition observed in the asymmetric case.

In the antisymmetric case, the transition is smooth (Fig. \ref{fig:fig2}F square symbols). At the transition value $\mu^*\approx 1.967$, the two equilibrium shapes U\textsubscript{A} and U\textsubscript{B} smoothly collapse onto each other and a single equilibrium configuration remains available thereafter. 
When approaching the transition, the shape of the strip varies sharply but in a continuous manner. For an infinitesimal variation $\delta \mu$, the variation of the mid-point deflection remains infinitesimal, which is also true for all points along the strip. This differs drastically from the two other types of actuation that lead to snap-through.

\section{Static equilibria in the Euler-Beam Model}
\label{sec:EB}

We carry out an analysis of the static equilibria of the strip under the three types of boundary actuation studied in Fig.~\ref{fig:fig2} in the context of the Euler-beam model.
Namely, we approximate the arclength $s$ by the $x$-coordinate for $x\in[-L/2,L/2]$, and we describe the deflection $w(x,t)$ by the linear Euler-beam equation~\cite{timoshenko2009}. Using the non-dimensional quantities \cite{gomez2017},
\begin{equation}		
W=\dfrac{w}{\sqrt{L\Delta L}},\qquad
X=\dfrac{x}{L},\qquad
T=\displaystyle{\sqrt{\frac{B}{\rho b h L^4}}t},
\label{eq:nonDimensionnalization}
\end{equation}
the linear Euler-beam equation \cite{timoshenko2009} takes the form
\begin{equation}
\frac{\partial^2 W}{\partial T^2}+\frac{\partial^4 W}{\partial X ^4}+\Lambda^2 \frac{\partial^2 W}{\partial X^2}=0,
\label{eq:beam_equation_nodim}
\end{equation}
where $\Lambda^2=F L^2/B$ is the non-dimensional longitudinal compression force. To close the model,  \eqref{eq:beam_equation_nodim} is complemented by a nonlinear incompressibility constraint that expresses the longitudinal confinement imposed to the beam by the boundaries \cite{pandey2014,gomez2017},
\begin{equation}
\int_{-1/2}^{1/2}\left(\frac{\partial W}{\partial X}\right)^2dX=2,
\label{eq:geometrical_constraint_nodim}
\end{equation}
and a set of four boundary conditions that depend on the three types of actuation.
Specifically, the boundary conditions at both ends $X=0,1$
of the strip are given in terms of the dimensionless parameter $\mu = \alpha\sqrt{L/\Delta L}$,
\begin{equation}
\begin{split}
&\left. W\right|_{X=0}= \left.W\right|_{X=1}= 0, \qquad \left.\frac{\partial W}{\partial X}\right|_{X=0}=
\mu,  \\
\textrm{asymmetric:} \left.\frac{\partial W}{\partial X}\right|_{X=1}=
0, &
\qquad 
\textrm{symmetric:} \left.\frac{\partial W}{\partial X}\right|_{X=1}= -\mu,
\qquad
\textrm{antisymmetric:} \left.\frac{\partial W}{\partial X}\right|_{X=1}= \mu.
\end{split}
\label{eq:BC}
\end{equation}

The static equilibria $W_\textrm{eq}(X)$ of the elastic strip are solutions of the steady counterpart of~\eqref{eq:beam_equation_nodim},
\begin{equation}
\frac{d^4 W_\textrm{eq}}{d X ^4}+\Lambda_\textrm{eq}^2 \frac{d^2 W_\textrm{eq}}{d X^2}=0,
\label{eq:beam_equation_nodim_static}
\end{equation}
whose general solution is of the form
\begin{equation}
W_\textrm{eq}(X)=A\sin(\Lambda_\textrm{eq} X)+B\cos(\Lambda_\textrm{eq} X)+ CX+D.
\label{eq:sol_general_static}
\end{equation}
Here, $A$, $B$, $C$, $D$ are 4 unknown constants that must be chosen so that \eqref{eq:sol_general_static} satisfies the appropriate boundary conditions. 
Writing the boundary conditions of the elastic strip yields a system of equations of the form,
	$\mathbf{M}\mathbf{v}=\mathbf{b}$,
where $\mathbf{v}=(A,B,C,D)$.  The geometric constraint
 \eqref{eq:geometrical_constraint_nodim} implies that the equilibrium configurations must also satisfy 
 \begin{equation}
 \int_{-1/2}^{1/2} \left(\dfrac{\partial W_\textrm{eq}}{\partial X}\right)^2 dX = 2.
 \label{eq:geometrical_constraint_equilibrium}
 \end{equation}
Together, the system of equations $\mathbf{M}\mathbf{v}=\mathbf{b}$  and \eqref{eq:geometrical_constraint_equilibrium} determine the eigenvalue $\Lambda_{\textrm{eq}}$ and eigenfunction $W_\textrm{eq}(X)$ by providing conditions to solve for $\Lambda_\textrm{eq}$ and $(A,B,C,D)$. Semi-analytic solutions are tabulated in the Supplemental Document of~\cite{radisson2022}. 
 
The homogeneous system $\mathbf{M}\mathbf{v}=\mathbf{0}$ corresponds to equilibrium states of the Euler-buckled strip. This case admits an infinite number of eigenvalues $\Lambda$ and corresponding eigenmodes $\mathbf{v}$, that come in pairs of increasing values of bending energy $\mathcal{E}_b$; the two smallest eigenvalues and associated eigenmodes correspond to the two first buckling modes U\textsubscript{A} and U\textsubscript{B}.
The two equilibria associated with the second mode are denoted by S\textsubscript{A} and S\textsubscript{B}, and those with the third mode by W\textsubscript{A} and W\textsubscript{B} (Fig.~\ref{fig:fig2Bis}, top row).  

Antisymmetric, symmetric, and asymmetric boundary actuation results in non-zero right-hand side $\mathbf{b}$,
 for which the eigenvalues $\Lambda$ and eigenvectors $\mathbf{v} = (A,B,C,D)$  are given in \cite{radisson2022}. For each boundary actuation, there exists an infinite number of eigenvalues and corresponding eigenmodes that describe how the equilibrium modes of the Euler-buckled strip are modified under the corresponding rotational actuation of the boundary. Each type of boundary actuation affects differently the equilibrium states of the strip.

For asymmetric boundary actuation, U\textsubscript{A}  monotonically approaches S\textsubscript{B} until they merge and both disappear at $\mu^* = 2$ (Fig.~\ref{fig:fig2Bis} and Fig.~\ref{fig:fig2}D).   
This value of $\mu^*$ corresponds to the threshold value where an abrupt snap-through transition is observed in our numerical simulations. 
For $\mu>\mu^*$, U\textsubscript{B} is the only available equilibrium. 
This analysis was carried out in \cite{gomez2017} but reviewed here to compare to the other cases.

Symmetric boundary actuation tends to symmetrize the two shapes S\textsubscript{A} and S\textsubscript{B} until they both merge with U\textsubscript{A} in a first bifurcation at $\mu^*=2$ (Fig. \ref{fig:fig2Bis} and \ref{fig:fig2}E). On the bifurcation diagram (Fig. \ref{fig:fig2}E), S\textsubscript{A} and S\textsubscript{B} are indistinguishable as, by symmetry of the system, they both have the same mid-point deflection (Fig. \ref{fig:fig2}E). However, plotting the same bifurcation diagram in term of the deflection at $X=-1/4$ (Fig. \ref{fig:fig2}H inset) confirms that they both collapse on U\textsubscript{A} at the exact same value of $\mu$ and that they approach the latter from both sides. For slightly larger values of $\mu$ the branch issued from these three branches merges with W\textsubscript{B} in a second bifurcation at $\mu_{2}^*\approx2.012$. At this bifurcation, the two branches -- one representing W\textsubscript{B} and the other representing the branch issued from the merging of U\textsubscript{A} with S\textsubscript{A} and S\textsubscript{B} -- disappear (Fig. \ref{fig:fig2}H). For larger values of $\mu$, U\textsubscript{B} is the only equilibrium available.  Our numerical simulations seem to indicate that the strip snaps to U\textsubscript{B} after reaching the first bifurcation at $\mu^*=2$, which confirms Gomez \textit{et al.} assertion that the strip snaps from an equilibrium that becomes unstable instead of from an equilibrium that suddenly disappears, as in the asymmetric case \cite{gomez2017}.

Antisymmetric actuation tends to anti-symmetrize the symmetric modes U\textsubscript{A} and U\textsubscript{B} until they both merge with S\textsubscript{B} at $\mu^*=2$ (Fig.~\ref{fig:fig2Bis} and Fig.~\ref{fig:fig2}F). As shown on the bifurcation diagram, for larger value of $\mu$ the branch issued from these three branches (U\textsubscript{A}, U\textsubscript{B},  and S\textsubscript{B}) remains the only equilibrium observed in the numerical simulations.

\section{Stability analysis}
\label{sec:stability}

We analyze the dynamics of the strip around the static equilibria using two approaches:  we carry a stability analysis based on the Euler-beam model, and we investigate the dynamics numerically using the discrete Cosserat rod model. 

Starting from the Euler beam model~(\ref{eq:beam_equation_nodim}-\ref{eq:geometrical_constraint_nodim}),
we consider the dynamic evolution of a small perturbation about the equilibrium state characterized by the shape of the strip $W_{\textrm{eq}}(X)$ and the compression force $\Lambda_{\textrm{eq}}$. For this purpose, we write the shape $W(X,T)$ and compression force $\Lambda(T)$  as follows \cite{pandey2014,nayfeh2008},
\begin{equation}
\left.
\begin{array}{c}		
W(X,T)=W_{\textrm{eq}}(X)+\epsilon W_{\textrm{p}}(X) e^{\sigma T},\quad \quad
\Lambda(T)=\Lambda_{\textrm{eq}}+\epsilon \Lambda_{\textrm{p}} e^{\sigma T}.
\end{array}
\right.
\label{eq:smallPerturbationExpression}
\end{equation}
Here, $W_{\textrm{p}}(X)$ is the shape of the perturbation, $\epsilon$ its amplitude (considered small $\epsilon \ll 1$), and $\sigma$ its growth rate. Substituting these expressions in \eqref{eq:beam_equation_nodim} and \eqref{eq:geometrical_constraint_nodim}, we get at first order in $\epsilon$,
\begin{equation}
\sigma^2 W_{\textrm{p}}+\frac{d^4 W_{\textrm{p}}}{d X^4}+\Lambda_{\textrm{eq}}^2\frac{d^2 W_{\textrm{p}}}{d X^2}=-2\Lambda_{\textrm{eq}}\Lambda_{\textrm{p}}\frac{d^2 W_{\textrm{eq}}}{d X^2}, \qquad \int_{-1/2}^{1/2}\frac{d W_{\textrm{eq}}}{dX}\frac{dW_{\textrm{p}}}{dX}dX=0.
\label{eq:smallperturbations_beameq}
\end{equation}
These equations describe the linear dynamic of the perturbation mode $W_{\textrm{p}}(X)$ around a given equilibrium configuration $W_{\textrm{eq}}(X)$. 
We require the general solution of the non-homogeneous ODE in \eqref{eq:smallperturbations_beameq} to satisfy the geometrical constraint in \eqref{eq:smallperturbations_beameq} and the appropriate boundary conditions. We obtain a nonlinear eigenvalue problem of the form $\mathbf{M}_\textrm{p}(\sigma^2)\mathbf{v}_\textrm{p}=0$, with eigenvalue $\sigma^2$ and eigenvector $\mathbf{v}_\textrm{p}$. 
This eigenvalue problem admits a solution only when $\det(\mathbf{M}_\textrm{p})=0$, which yields a nonlinear equation of $\sigma^2$ that cannot be solved analytically.  To obtain the eigenvalues $\sigma^2$  associated with each equilibrium configuration, we numerically search for eigenvalues in a specific domain (here, we checked in the range $0<|\sigma^2|<50000$), thus excluding all eigenvalues that are beyond this range.

For $\mu=0$, for the two first equilibria U\textsubscript{A} and U\textsubscript{B}, we find only negative eigenvalues, indicating that these fundamental equilibria are stable. This result agrees with our numerical simulations based on the Cosserat rod theory.  
The remaining equilibria (S\textsubscript{A}, S\textsubscript{B}, W\textsubscript{A}, W\textsubscript{B}, etc.) possess at least one positive eigenvalue, confirming that these modes are unstable. This explains why they are not observed in forward-time numerical simulations. 

\begin{figure}[t]
 	\centering
 	\includegraphics[width =\textwidth]{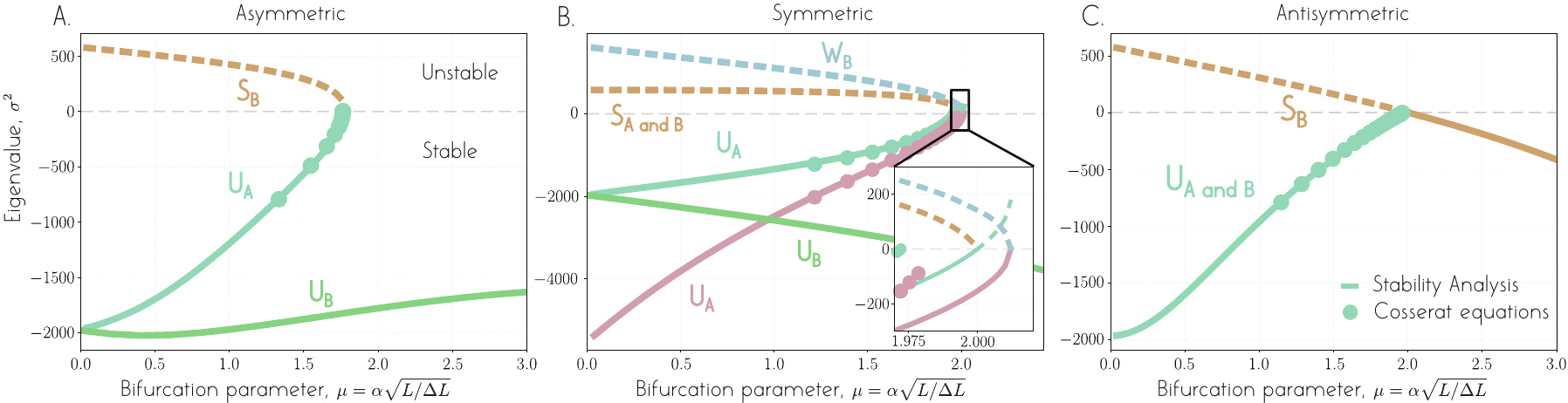}
 	\caption{\footnotesize{\textbf{Linear stability analysis} gives access to the eigenvalues associated with the modes of perturbation of the static equilibria. For each equilibrium configuration, we show the evolution in terms of $\mu$ of the eigenvalue that has the smallest absolute value (fundamental mode). Data obtained from  numerical analysis of the impulse response of the strip (Fig.~\ref{fig3}) are shown as dotted symbols for comparison. In the Symmetric case, for the U\textsubscript{A} equilibrium, both the lowest eigenvalue for U\textsubscript{A} (fundamental mode) and the second lowest eigenvalue U\textsubscript{A} (first harmonic) are shown.}}
 	\label{fig:fig4}
 \end{figure}

We next examine the evolution of the eigenvalues $\sigma^2$ associated with U\textsubscript{A,B}, S\textsubscript{A,B}, and W\textsubscript{A,B} when the strip boundaries are rotated (Fig. \ref{fig:fig4}). 
For each equilibrium shape, we focus on the fundamental eigenvalue that possesses the smallest absolute value. 

For the asymmetric actuation (Fig.~\ref{fig:fig4}A), 
the equilibrium configuration U\textsubscript{B} remains stable for all values of $\mu$ considered, whereas the fundamental eigenvalue of U\textsubscript{A} monotonically increases until it reaches the zero axis and disappears. Indeed, as $\mu$ increases, the left boundary is rotated  to the opposite side compared to the buckling side of the stable U\textsubscript{A} configuration, and this equilibrium becomes less favorable, and thus less stable. Simultaneously, the eigenvalue of S\textsubscript{B} monotonically decreases until its positive eigenvalue hits the zero axis at the exact location ($\mu^*\approx1.7812$) where U\textsubscript{A} disappears. This is indicative of a saddle-node bifurcation as demonstrated in \cite{gomez2017}.

For the symmetric actuation (Fig.~\ref{fig:fig4}B),   
as $\mu$ increases, both boundaries are rotated towards the buckling side of U\textsubscript{B}. This equilibrium becomes more favorable, which explains the monotonic decrease of the eigenvalue observed for this equilibrium in Fig. \ref{fig:fig4}B. Meanwhile, U\textsubscript{A} has to bend more to satisfy the boundary conditions and becomes less stable as $\mu$ increases. This is reflected by the monotonic increase of the associated eigenvalues. The fundamental mode of perturbation becomes unstable at $\mu^*=2$ where its eigenvalue (labelled U\textsubscript{A0}) crosses zero (inset in Fig. \ref{fig:fig4}B).  Meanwhile, the  fundamental eigenvalues of the two S shapes remain equal to each other for all values of $\mu$; they slowly decrease until they reach zero at $\mu^*=2$ where they both disappear. That is, at $\mu^*=2$, the two unstable solutions S\textsubscript{A} and S\textsubscript{B} collapse onto the stable solution U\textsubscript{A} and disappear while the latter becomes unstable; this is typical of a subcritical pitchfork bifurcation.

Interestingly, a second bifurcation occurs at $\mu_{2}^*\approx2.012$, the eigenvalue (labelled U\textsubscript{A1}) corresponding to the first harmonic mode of perturbation around U\textsubscript{A} (i.e., the mode associated with the eigenvalue that has the second smallest absolute value) and the one associated with the fundamental mode of perturbation around W\textsubscript{B} both vanish. At this point, the two corresponding equilibria (U\textsubscript{A} and W\textsubscript{B}) suddenly disappear. 
That is, at $\mu_{2}^*\approx2.012$, 
an unstable mode of perturbation associated with W\textsubscript{B} and a stable mode associated with U\textsubscript{A} collapse and suddenly disappear. This is typical of a saddle-node bifurcation.

For the antisymmetric case (Fig. \ref{fig:fig4}C), as $\mu$ increases, the eigenvalues associated with U\textsubscript{A} and U\textsubscript{B} evolve in the same way. The eigenvalues increase monotonically until they reach zero at $\mu^*=2$ where they disappear. Meanwhile, the eigenvalues of S\textsubscript{A} and S\textsubscript{B} evolve in the opposite way. The eigenvalue of S\textsubscript{A} (not plotted on the figure) monotonically increases (becoming more unstable) with increasing $\mu$ while S\textsubscript{B} monotonically decreases. This is due to the direction of rotation of the boundaries which makes S\textsubscript{A} (respectively S\textsubscript{B}) a less favorable (respectively more favorable) state as $\mu$ increases. The eigenvalue of S\textsubscript{B} decreases until reaching zero at $\mu^*=2$, beyond which it becomes negative causing the corresponding mode to switch from unstable to stable. 
Thus, at $\mu^*=2$, the two stable equilibria U\textsubscript{A} and U\textsubscript{B} collapse on the unstable equilibrium S\textsubscript{B} and disappear while the latter becomes stable, this is indicative of a supercritical pitchfork bifurcation.

\begin{figure}[t]
	\centering
	\includegraphics[width =\textwidth]{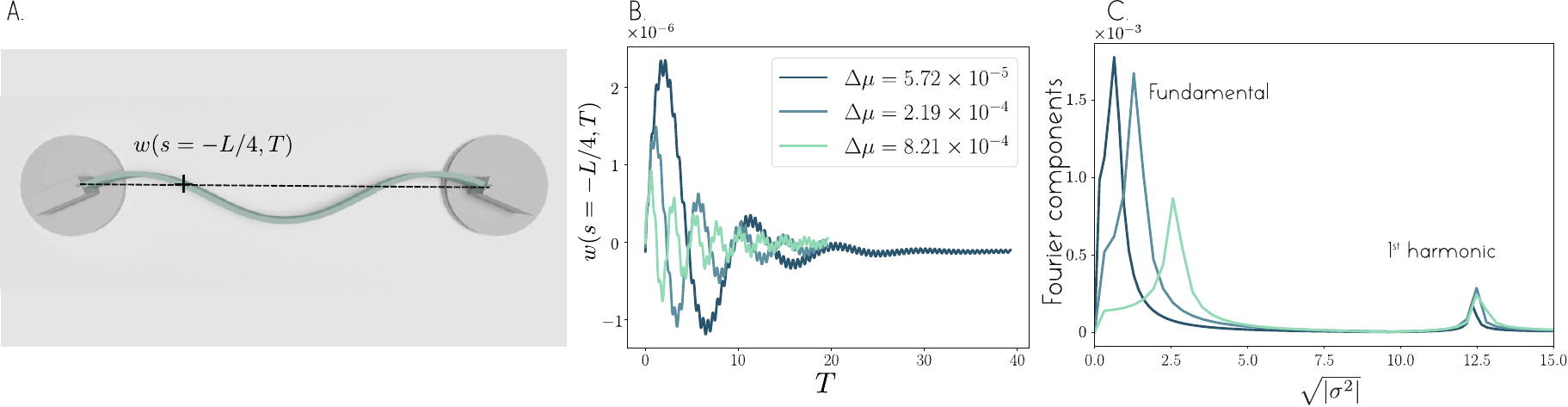}
	\caption{\footnotesize{\textbf{Linear dynamics of the strip around the equilibrium configurations.} (A.) From a strip that remains at its equilibrium configuration, we apply an instantaneous point force on a vertex of the Cosserat rod in the transverse direction $\mathbf{e}_y$.
 Starting from the U\textsubscript{A} equilibrium, we apply an initial kick at the longitudinal coordinate $s=-L/4$.
 (B.) We then obtain the impulse response of the strip by recording the transverse position at the longitudinal coordinate $s=-L/4$ after the initial kick. (C.) The eigenpulsations $\sqrt{|\sigma^2|}$ associated with the vibrations of the strip are obtained by performing a Fourier transform of this signal. The process is repeated for different equilibrium configurations at different distances $\Delta\mu=\mu-\mu^*$ from the bifurcation point. Here, the  procedure is shown for the Symmetric case but is repeated in the exact same way for the two other configurations.}}
	\label{fig:fig3}
\end{figure}

To complete this analysis, we probe the linear dynamics of the strip around its equilibria numerically using the Cosserat rod theory. 
The general process is illustrated in Fig. \ref{fig:fig3} in the case of the symmetric boundary actuation.
We first hold the strip in a stable configuration, with the value of $\mu$ held fixed.  
At $t=0$, we impose a sudden kick to the strip by applying an instantaneous point force in the transverse direction at a vertex of the Cosserat rod in order to obtain its impulse response (Fig. \ref{fig:fig3}A). 
The magnitude of this force is chosen such that the amplitude of the subsequent oscillations remains small.
Following this kick, the strip is left free to oscillate (Fig. \ref{fig:fig3}B),
and its response is analyzed by performing a Fourier transform of the signal obtained from measuring the vertical position of one vertex of the strip against time (Fig. \ref{fig:fig3}C). The associated frequencies  reflect the eigenfrequencies $\sqrt{|\sigma^2|}$ of  the perturbation;  In Fig. \ref{fig:fig3}C, we show the fundamental mode and the first harmonic of perturbation only.
The process is repeated for different values of $\mu$, corresponding to various distances $\Delta\mu=\mu-\mu^*$ from the bifurcation point. 
 The eigenfrequencies decrease   with decreasing $\Delta \mu$ as the system gets closer to the bifurcation point. This `slowing down' phenomenon, studied in \cite{gomez2017} for the asymmetric actuation, is typical of systems that are approaching a transition \cite{scheffer2009}. 
 
We repeated this procedure for all three types of boundary actuation. The frequencies obtained from this vibration analysis are superimposed onto~\ref{fig:fig4} (colored markers). These data points are quantitatively consistent with the data obtained from the linear stability analysis (solid lines) except in the very vicinity of the bifurcation where the frequencies obtained numerically hit the zero axis before the ones obtained analytically (inset in Fig. \ref{fig:fig3}C). Indeed, the quasi-linear Euler beam model overestimate the value of the bifurcation point, which can also be observed from the static analysis reported in the previous section (Fig. \ref{fig:fig2}D-F).

Taken together, our stability analysis and numerical investigation reveal the stability of the static equilibria of the strip and indicate the type of bifurcation that occurs near shape transitions. 
In order to confirm the nature of the bifurcation and get a better understanding of the strip behavior in the vicinity of these transitions, we perform an asymptotic analysis near the bifurcation for each type of boundary actuation as discussed next.


\section{Asymptotic analysis}
\label{sec:asymptoticAnalysis}

When a strip starts from rest in a configuration that is close to its equilibrium configuration at the bifurcation point $\mu^\ast$, we expect its dynamic evolution to be slow due to the critical slowing down
of dynamical systems 
near a bifurcation \cite{strogatz1994,scheffer2009,gomez2017}. This is evident from the linear dynamic analysis in \S\ref{sec:stability}; as the eigenvalues vanish at the bifurcation point, the typical time scale associated with the corresponding modes diverges to infinity. To capture this slowing down, we introduce a slow time $\tau=\Delta \mu^{a} T$ (see \cite{gomez2017}). To describe the dynamics of the strip in the vicinity of the bifurcation, we expand its state $(W(X,\tau), \Lambda(\tau))$ at a given time $\tau$ in terms of powers of $\Delta \mu$ as follows,
\begin{equation}
\begin{split}		
W(X,\tau)&= W_{\textrm{eq}}^*(X)+\Delta\mu^{b}W_0(X,\tau)+\Delta\mu^{{b}_1} W_1(X,\tau)+ \text{h.o.t.},
\\
\Lambda(\tau)&=\Lambda_{\textrm{eq}}^*+\Delta\mu^{c}\Lambda_0(\tau)+\Delta\mu^{{c}_1} \Lambda_1(\tau)+ \text{h.o.t},
\end{split}
\label{eq:expansion}
\end{equation}
where typically ${b}_1=\min(2b, b+c)$ and ${c}_1=\min(2c, b+c)$ and so on for higher-order terms.

 The values of the three parameters $a$, $b$, and $c$ depend on the intrinsic properties of the system. In \cite{gomez2017}, $a$ was set to $a=1/4$ and it was postulated that $b=c=1/2$. In \cite{gomez2018b}, the same values for $a$, $b$ and $c$ were found starting from the assumptions that $a>0$ and $b=c \in \,  ]0,1[$. Here, we show that the values of $a$, $b$ and $c$, can be obtained, for all three set of boundary conditions, by exploiting the results of our static analysis and linear dynamic analysis reported in the previous sections.
In particular, we find that the values postulated in \cite{gomez2017} for the asymmetric case are correct but are different from the values obtained for the symmetric and antisymmetric cases, where
 we find that $b \neq c$, in contrast to the main assumption in \cite{gomez2017, gomez2018b}.

\begin{figure}[!t]
 	\centering
 	\includegraphics[width =\textwidth]{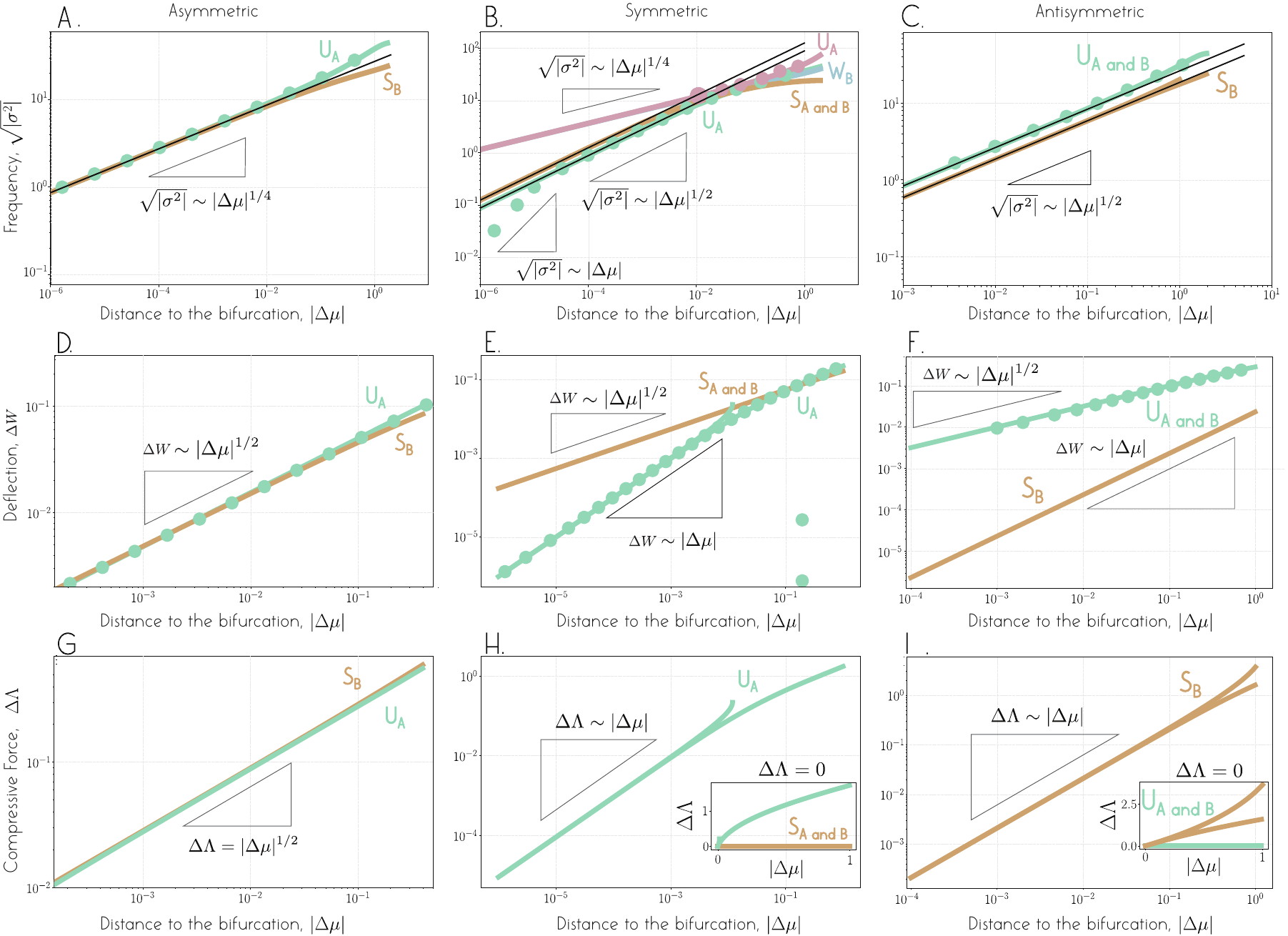}
 	\caption{\footnotesize{\textbf{Scaling of the different variables in the vicinity of the shape transition} When the system is pulled away from the bifurcation point, the different variables of the problem go away from their values at the bifurcation following a characteristic scaling. Here, for each configuration, we plot: (A, B, C) the eigenpulsations/growth rates, (D, E, F) the quantity $\Delta w$ and (G, H, I) the quantity $\Delta \Lambda$, against the distance to the bifurcation $|\Delta \mu|$ on a logarithmic scale. The data obtained from the static and stability analysis (full lines) are compared to the numerical data (dotted markers). For the equilibria that exist on both sides of the bifurcation, two branches are visible, each of them corresponding to one side of the bifurcation ($\Delta \mu<0$ and $\Delta \mu>0$).}}
 	\label{fig:fig5}
 \end{figure}

\subsection{Slow Time Scale Near the Bifurcation}
\label{sec:scalingsTime}

Our goal is to determine the scaling laws that govern the behavior of the strip near each bifurcation as a function of the perturbation $\Delta\mu=\mu-\mu^*$ away from the bifurcation.  

We first determine the value of $a$, which characterizes how the typical time scale of the system slows down when approaching the bifurcation. This typical time scale is simply related to the value of $\sigma$ 
associated with the critical modes. 
In Fig.~\ref{fig:fig5}A-C, using the data from Fig.~\ref{fig:fig4}, we plot
$\sqrt{|\sigma^2|}$ associated with each of these modes as a function of the distance $|\Delta \mu|$ from the bifurcation. 
Because the Cosserat and Euler-beam models lead to different $\mu^*$ values (they converge in the limit $\Delta L \rightarrow 0$), the values of $\Delta\mu=\mu-\mu^*$ are calculated using the corresponding bifurcation value (Table~\ref{tab:muStarNumRot}).
The results are shown on a logarithmic scale. 
In the very vicinity of the bifurcation ($\Delta\mu\ll 1$), the slopes reveal the slowing down exponent $a$. Clearly, the typical time scale associated with the dynamics of the strip diverges when approaching the bifurcation for all three boundary actuation. However, details of the slowing down vary depending on the type of boundary actuation. 

For the asymmetric case (Fig. \ref{fig:fig5}A), the typical time scale diverges as $T \sim\Delta\mu^{-1/4}$, thus $a=1/4$ as proposed in \cite{gomez2017}. In \cite{gomez2017}, this result was confirmed experimentally by measuring the typical time of snapping as the system is pulled to the right of the bifurcation, which we reproduce numerically using the Cosserat model in Appendix~\ref{sec:snappingTime}. The results in Fig. \ref{fig:fig5}A provide an analytical justification for the choice $a=1/4$ and confirm the robustness of the empirical observations in~\cite{gomez2017} by demonstrating the existence of the same scaling law to the left of the bifurcation.

For the symmetric case (Fig. \ref{fig:fig5}B), although the three equilibrium shapes U\textsubscript{A} and S\textsubscript{A,B} that interact at the first bifurcation have different $\sqrt{|\sigma^2|}$ values, the scaling is the same for all three. This scaling corresponds to $a=1/2$. Except very close to the bifurcation ($\Delta\mu\ll1$) where the Cosserat data (green markers) exhibit a stronger slowing down (see discussion in appendix \ref{sec:overDampedBoundaryLayer}). In Fig. \ref{fig:fig5}B, we also plot the eigenfrequency of the first harmonic (second eigenvalue) of U\textsubscript{A} and fundamental mode (first eigenvalue) associated with W\textsubscript{B}. These two modes interact through a secondary bifurcation and the slowing down associated with it follows the scaling $a=1/4$. 
Our asymptotic analysis is concerned with the primary bifurcation for which $a = 1/2$.

For the antisymmetric case (Fig. \ref{fig:fig5}C), the typical time scale associated with the fundamental mode of perturbation of U\textsubscript{A}, U\textsubscript{B}, and S\textsubscript{B} follows the same scaling with $a=1/2$.

\subsection{Asymptotic Expansion Near the Bifurcation}
\label{sec:scalings}

To estimate the values of $b$ and $c$ in the asymptotic expansions in~\eqref{eq:expansion}, which characterize how quickly the shape $W(X)$ and compression force $\Lambda$ move away from their respective values $W_{\textrm{eq}}^*(X)$ and $\Lambda_{\textrm{eq}}^*$ at the bifurcation, we plot, as a function of $\Delta \mu$,  the $L^1$ norms $\Delta W$ (Fig.~\ref{fig:fig5}D-F) and  $\Delta \Lambda$ (Fig.\ref{fig:fig5}G-I) for the equilibria involved at the bifurcation,
\begin{equation}
\Delta W=\int_{0}^{1}\sqrt{\left(W_{\textrm{eq}}(X, \Delta\mu)-W_{\textrm{eq}}^*(X)\right)^2}dX, \qquad
\Delta \Lambda=\sqrt{(\Lambda_{\textrm{eq}}(\Delta\mu)-\Lambda_{\textrm{eq}}^*)^2}.
\label{eq:deltaW}
\end{equation}
Close to the bifurcation, $\Delta W$ and $\Delta \Lambda$ provide estimates of the amplitude of the leading order mode  in \eqref{eq:expansion}.

Clearly, $\Delta W$ and $\Delta \Lambda$ scale differently with $\Delta \mu$ based on the type of boundary actuation. 
For the asymmetric case, the two equilibria U\textsubscript{A} and S\textsubscript{B} follow the same scalings $\Delta W \propto|\Delta\mu|^{1/2}$ and $\Delta \Lambda \propto|\Delta\mu|^{1/2}$. This imply that there is only one  route (one mode) available to go away from the bifurcation point. This analysis provides an analytic justification for the choice $b=1/2$ and $c=1/2$ adopted by Gomez \textit{et al.} \cite{gomez2017}.

For the symmetric and antisymmetric cases, 
there are two different routes available to go away from the bifurcation:  {following either the S\textsubscript{A,B} branch or U\textsubscript{A,B} branch. In the symmetric case,  the system goes away from the bifurcation following the U\textsubscript{A} branch for $\Delta \mu >0$  and S\textsubscript{A,B} for $\Delta \mu <0$.
In the antisymmetric case,  the system goes away from the bifurcation following the S\textsubscript{B} branch for $\Delta \mu >0$  and U\textsubscript{A,B} for $\Delta \mu <0$.}
This leads to the existence of two different forms of scaling:
{in the symmetric case, 
following U\textsubscript{A}
results in
$\Delta W\propto|\Delta \mu|$ and $\Delta \Lambda\propto|\Delta \mu|$ while following  
S\textsubscript{A,B} results in
$\Delta W\propto|\Delta \mu|^{1/2}$ and $\Delta \Lambda=0$ (Fig.\ref{fig:fig5}E and inset in Fig. \ref{fig:fig5}H);
in the antisymmetric case, 
following S\textsubscript{B} 
results in $\Delta W\propto|\Delta \mu|$ and $\Delta \Lambda\propto|\Delta \mu|$
while following U\textsubscript{A,B} results in
$\Delta W\propto|\Delta \mu|^{1/2}$ and $\Delta \Lambda=0$ (Fig.\ref{fig:fig5}F and inset in Fig. \ref{fig:fig5}I).}
 This suggests two different expansions of $W(X,\tau)$ and $\Lambda(\tau)$ depending on the route the strip takes to move away from the bifurcation: in one route, $W(X,\tau)$ would be expanded in powers of $\Delta \mu^{1/2}$ and $\Lambda$ would be constant, whereas in the other route, $W(X,\tau)$ and $\Lambda$ would be expanded in powers of $\Delta \mu$. In general, 
the solution could follow any linear combination of these two routes. Thus, the general expansion of $W(X,\tau)$ and $\Lambda(\tau)$ in terms of $\Delta \mu$ would be associated with $b=1/2$ and $c=1$ with subsequent exponents that satisfy 
\begin{equation}
\begin{split}		
W(X,\tau)& = W_{\textrm{eq}}^*(X)+\Delta\mu^{1/2} W_{0}(X,\tau)+\Delta\mu W_{1}(X,\tau)+O(\Delta \mu^{3/2}),\\
\Lambda(\tau)&=\Lambda_{\textrm{eq}}^*+\Delta\mu\Lambda_{0}(\tau)+O(\Delta \mu^{3/2}). 
\end{split} 
\label{eq:expansionPF}
\end{equation}


\subsection{Asymptotic analysis}\label{sec:asymptoticAnalysisSubsection}

We substitute $\tau=\Delta\mu^{a}T$ in the system of equations (\ref{eq:beam_equation_nodim},\ref{eq:geometrical_constraint_nodim},\ref{eq:BC}) and simplify to arrive at
\begin{equation}
\begin{split}
& \Delta \mu^{2a}\frac{\partial^2W}{\partial \tau^2}+\frac{\partial^4 W}{\partial X^4}+\Lambda^2\frac{\partial^2 W}{\partial X^2}=0, \qquad \int_{0}^{1} \left(\dfrac{\partial W}{\partial X}\right)^2dX=2, 
 \\[3mm]
& \qquad \left. W\right|_{X=0}= \left. W\right|_{X=1}=0,\quad  \ \left. \dfrac{\partial W}{\partial X}\right|_{X=0}=\mu^*+\Delta \mu, 
 \\[3mm] 
\text{asymmetric:} \ \left. \dfrac{\partial W}{\partial X}\right|_{X=1}&=0, 
\qquad
\text{symmetric:} \ \left. \dfrac{\partial W}{\partial X}\right|_{X=1} =-\mu^*-\Delta\mu,
\qquad 
\text{antisymmetric} \ \left. \dfrac{\partial W}{\partial X}\right|_{X=1} =\mu^*+\Delta\mu. 
\label{eq:beam_slow_bifurcation}
\end{split}
\end{equation}

We next substitute the expansion \eqref{eq:expansion} into~\eqref{eq:beam_slow_bifurcation}, with the appropriate exponents for each type of actuation, and write the leading order mode  $(W_0(X,\tau), \Lambda_{0}(\tau))$
in terms of its
shape $\Phi_0(X)$ and amplitude $\mathcal{A}(\tau)$,
\begin{equation}
    W_0(X,\tau) = \mathcal{A}(\tau)\Phi_0(X), \qquad  \Lambda_{0}(\tau)=\mathcal{A}(\tau).
    \label{eq:leadingmode}
\end{equation}

For the asymmetric case, with the scaling $a=1/4$, $b=1/2$ and $c=1/2$, the problem is solved by Gomez \textit{et al.} \cite{gomez2017} (see Appendix~\ref{sec:asymptotic_asym}). 
Here, we perform the same analysis for the symmetric and antisymmetric cases, which have the same scaling $a=1/2$, $b=1/2$ and $c=1$. The system obtained for these two cases is the same and only the boundary conditions are different.
Introducing the linear operator $\mathcal{L}$,
\begin{equation}
\mathcal{L}  (\cdot)=\dfrac{\partial^4 (\cdot)}{\partial X^4}+(\Lambda_{\textrm{eq}}^*)^2\dfrac{\partial^2(\cdot)}{\partial X^2}, \qquad
\label{eq:L}
\end{equation}
we get, at the leading order $O(\Delta \mu^{1/2})$, 
\begin{equation}
\begin{split}
\mathcal{L}(\Phi_0) &  = 0,
\qquad 
\int_{0}^{1}\frac{d W_{\textrm{eq}}^*}{d X}\frac{d \Phi_0}{dX}dX=0,  \\
 \left. \Phi_0\right|_{X=0} = \left. \Phi_0\right|_{X=1} & =0,\qquad  \ \left. \dfrac{\partial \Phi_0}{\partial X}\right|_{X=0}=0, \qquad
\left. \dfrac{\partial \Phi_0}{\partial X}\right|_{X=1}=0.
\label{eq:O1}
\end{split}
\end{equation}
This leading order system is
homogeneous with homogeneous boundary conditions. The solution to \eqref{eq:O1} provides an expression for the eigenmode $\Phi_0(X)$ of the form
$\Phi_0(X)=A_0\sin(\Lambda_{\textrm{eq}}^* X)+B_0\cos(\Lambda_{\textrm{eq}}^* X)+C_0X+D_0$,
where $A_0$, $B_0$, $C_0$ and $D_0$ are determined from boundary conditions. Using the corresponding expression for $W_{\textrm{eq}}^*(X)$, we arrive at
\begin{equation}
\begin{split}
\textrm{symmetric:} & \ \Phi_0(X)=\sin\left(\Lambda_{\textrm{eq}}^* X\right)-\Lambda_{\textrm{eq}}^* X+\frac{\Lambda_{\textrm{eq}}^*-\sin(\Lambda_{\textrm{eq}}^*)}{\cos(\Lambda_{\textrm{eq}}^*)-1}\left(\cos(\Lambda_{\textrm{eq}}^* X)-1\right), \\
\textrm{antisymmetric:} & \ \Phi_0(X)=\cos(\Lambda_{\textrm{eq}}^* X)-1. \\
\end{split}
\label{eq:leadingOrderSolPF}
\end{equation}

At $O(\Delta\mu)$, we get the system
\begin{equation}
\begin{split}
 \mathcal{L}(W_1) & = -2\Lambda_{\textrm{eq}}^*\Lambda_0\frac{d^2 W_{\textrm{eq}}^*}{d X^2},
\qquad 
 \int_{0}^{1}\frac{d W_{\textrm{eq}}^*}{d X}\frac{\partial W_1}{\partial X}dX=-\frac{1}{2}\mathcal{A}^2 \int_{0}^{1}\!\!\left(\dfrac{d\Phi_0}{dX}\right)^2\!dX,  
\\
 \left. W_1\right|_{X=0} = \left. W_1\right|_{X=1} & =0, \qquad   
\left. \dfrac{\partial W_1}{\partial X}\right|_{X=0}  =0, 
\quad 
\textrm{symmetric:}  \ \left. \dfrac{\partial W_1}{\partial X}\right|_{X=1}=-1, \quad \textrm{antisymmetric:} \ \left. \dfrac{\partial W_1}{\partial X}\right|_{X=1}=1.
\label{eq:O2}
\end{split}
\end{equation}
At this order, the system is non-homogeneous with non-homogeneous boundary conditions, but contrary to the asymmetric case (see \cite{gomez2017}), it is independent of time because of the higher value of $a$ in the symmetric and antisymmetric actuation. The time derivative comes into play only at next order. The solution of \eqref{eq:O2} is of the form
\begin{equation}
W_1(X)=A_1\left\{\sin(\Lambda_{\textrm{eq}}^* X)-\Lambda_{\textrm{eq}}^* X \right\}+B_1\left\{\cos(\Lambda_{\textrm{eq}}^* X)-1 \right\}+X\left\{1+\frac{\Lambda_0}{\Lambda_{\textrm{eq}}^*}\left(\frac{dW_{\textrm{eq}}^*}{dX}-\mu^*\right)\right\}.
\label{eq:secondOrderSolPitchfork}
\end{equation}
To obtain this form of the solution, we used the two boundary conditions at $X=0$. Expressions for $A_1$, $B_1$ and $\Lambda_{0}$ must be determined from the two remaining boundary conditions and the geometrical constraint. For the symmetric case, we get
\begin{equation}
A_1=\displaystyle \frac{\Lambda_{\textrm{eq}}^*-\Lambda_{0}\mu^*\left(1+\cos(\Lambda_{\textrm{eq}}^*)\right)}{(\Lambda_{\textrm{eq}}^*)^2},\qquad
B_1=\displaystyle \frac{2\Lambda_0\mu^*-\Lambda_{\textrm{eq}}^*+A_1\Lambda_{\textrm{eq}}^*\left(\Lambda_{\textrm{eq}}^*-\sin(\Lambda_{\textrm{eq}}^*)\right)}{\Lambda_{\textrm{eq}}^*\left(\cos(\Lambda_{\textrm{eq}}^*)-1\right)},\qquad
\Lambda_0=\displaystyle \mathcal{C}_1+\mathcal{C}_2\mathcal{A}^2,
\label{eq:constantsSecondOrderSymm}
\end{equation} 
where $\mathcal{C}_1$ and $\mathcal{C}_2$ are two constants given by
\begin{equation}
\begin{split}
\mathcal{C}_1&=\displaystyle\frac{4\Lambda_{\textrm{eq}}^*}{2\mu^*}, 
\\
\mathcal{C}_2&=\displaystyle\frac{(\Lambda_{\textrm{eq}}^*)^2\left\{2(\Lambda_{\textrm{eq}}^*)^3+4\Lambda_{\textrm{eq}}^*\left[\cos(\Lambda_{\textrm{eq}}^*)-\cos(2\Lambda_{\textrm{eq}}^*)-\Lambda_{\textrm{eq}}^*\sin(\Lambda_{\textrm{eq}}^*)\right]-(\Lambda_{\textrm{eq}}^*)^2\sin(2\Lambda_{\textrm{eq}}^*)+2\sin(\Lambda_{\textrm{eq}}^*)-4\sin(\Lambda_{\textrm{eq}}^*)\right\}}{8(\mu^*)^2\sin^4(\Lambda_{\textrm{eq}}^*/2)}. 
\label{eq:constantsLambda0Symm}
\end{split}
\end{equation} 
For the antisymmetric case, we get
\begin{equation}
A_1= \frac{1}{\Lambda_{\textrm{eq}}^*}, \qquad B_1=0, \qquad
\Lambda_0= -\frac{2\Lambda_{\textrm{eq}}^*}{3(\mu^*)^2}\left(\frac{\mathcal{A}^2(\Lambda_{\textrm{eq}}^*)^2}{2}+\mu^*\right).
\label{eq:constantsSecondOrderAntisymm}
\end{equation} 

At order $O(\Delta\mu^{3/2})$, the system of equations is given by
\begin{equation}
\begin{split}
&\mathcal{L}(W_2)= -2\Lambda_{\textrm{eq}}^*\Lambda_0\mathcal{A}\frac{d^2 \Phi_0}{d X^2}-\Phi_0\frac{d^2\mathcal{A}}{d\tau^2}, \qquad
 \int_{0}^{1}\frac{d W_{\textrm{eq}}^*}{d X}\frac{\partial W_2}{\partial X}dX=- \int_{0}^{1}\left(\dfrac{\partial \Phi_0}{\partial X}\right)\left(\dfrac{\partial W_1}{\partial X}\right)dX, \\  
&\qquad \left. W_2\right|_{X=0}= \left. W_2\right|_{X=1}=0, \qquad   
\left. \dfrac{\partial W_2}{\partial X}\right|_{X=0}  =0,  \qquad \left. \dfrac{\partial W_2}{\partial X}\right|_{X=1}=0.
\label{eq:O3}
\end{split}
\end{equation}
The second order time derivative of the leading order mode comes into play on the right-hand side of the PDE in \eqref{eq:O3}. Following the same procedure as in \cite{gomez2017}, we now seek a solvability condition for \eqref{eq:O3} by requiring the non-homogeneous right hand side to be orthogonal to the adjoint solution. As with the operator for the asymmetric case, $\mathcal{L}$ is self-adjoint relative to the standard Cartesian scalar product and the adjoint solution is simply $\Phi_0(X)$. The resulting solvability condition takes the form 
\begin{equation}
\frac{d^2\mathcal{A}}{d\tau^2}=\frac{2\Lambda_{\textrm{eq}}^*\Lambda_0 I_2}{I_1}\mathcal{A},
\label{eq:solvabilityCond}
\end{equation}
where $I_1=\int_{0}^{1}\Phi_0^2dX$ and $I_2=\int_{0}^{1}(d\Phi_0/dX)^2dX$. 
Substituting the expression of $\Lambda_0$ yields an expression of the form
\begin{equation}
	\dfrac{d^2 \mathcal{A}}{d\tau^2}= b_{1,(\cdot)}  \mathcal{A}+b_{2, (\cdot)} \mathcal{A}^3,
	\label{eq:solvabilityPitchfork}
\end{equation}
where, for the symmetric case, $b_{1,\textrm{sym}} $ and $b_{2,\textrm{sym}} $ are two positive constants defined by
\begin{equation}
b_{1,\textrm{sym}} =\frac{2\Lambda_{\textrm{eq}}^* I_2\mathcal{C}_1}{I_1}, \qquad b_{2,\textrm{sym}} =\frac{2\Lambda_{\textrm{eq}}^* I_2\mathcal{C}_2}{I_1},
\label{eq:c1Andc2}
\end{equation}
and for the antisymmetric case, $b_{1,\textrm{antisym}} $ and $b_{2,\textrm{antisym}} $ are two negative constants defined by

\begin{equation}
b_{1,\textrm{antisym}} =-\frac{4(\Lambda_{\textrm{eq}}^*)^4}{9\mu^*}, \qquad b_{2,\textrm{antisym}} =-\frac{2(\Lambda_{\textrm{eq}}^*)^6}{9(\mu^*)^2}.
\label{eq:b1Andb2}
\end{equation}

Equation \eqref{eq:solvabilityPitchfork} describes the dynamics of the amplitude of the leading order mode. For both symmetric and antisymmetric actuation, the reduced form has the same functional form with one linear term and one cubic term (in contrast to one constant term and one square term for the amplitude equation obtained by Gomez \textit{et al.} for the asymmetric case). For the symmetric and antisymmetric cases, only the sign and values of the multiplicative constants in front of the linear and cubic terms are different. The different forms of the amplitude equations reflect the different kinds of bifurcation the three systems undergo at the shape transition.

\section{Analysis of the amplitude equations}
\label{sec:Aeq}

The solvability conditions~\eqref{eq:solvCondAsym} and~\eqref{eq:solvabilityPitchfork}
in the very vicinity of the bifurcation under asymmetric, symmetric and antisymmetric actuation are dynamical equations that describe the slow-time evolution of $\mathcal{A}(\tau)$ of the leading order mode $\Phi_0(X)$ at the bifurcation. In this section, we introduce the unscaled amplitude $A(T)$ of the leading order mode $\Phi_0(X)$ which characterizes how the strip goes away from its bifurcation shape in real time $T$ as opposed in rescaled time $\tau$. We then analyze the resulting amplitude equations to highlight the dynamical features in the vicinity of the elastic shape transitions under asymmetric, symmetric, and antisymmetric actuation.   

\subsection{Amplitude equations}
\label{sec:amplitude}

We substitute~\eqref{eq:leadingmode} back into \eqref{eq:expansionSN} and \eqref{eq:expansionPF}, and neglect terms of order $O(\Delta\mu)$ and higher. We get that the amplitude $\mathcal{A}(\tau)$ can be obtained directly from the asymptotic expansion of $W(X,\tau)$ such that
$\mathcal{A}(\tau)=\Delta\mu^{-1/2}{(W(X,\tau)-W_{\textrm{eq}}^*(X))}/{\Phi_0(X)}.$
By analogy, we introduce the unscaled amplitude $A(T)$ of the leading order mode $\Phi_0(X)$, which is related to $\mathcal{A}$ via the following scaling
\begin{equation}
A(T)\equiv\frac{W(X,T)-W_{\textrm{eq}}^*(X)}{\Phi_0(X)}=\Delta\mu^{1/2}\mathcal{A}(\tau=\Delta \mu^a T). 
\label{eq:actualAmplitude}
\end{equation}

In the case of asymmetric actuation, we substitute the amplitude equation~\eqref{eq:solvCondAsym} into~\eqref{eq:actualAmplitude} to arrive at the evolution equations of the amplitude $A(T)$,
\begin{equation}
\frac{d^2 A}{dT^2}=a_1\Delta\mu+a_2 A^2,
\label{eq:saddleNodeFinalForm}
\end{equation}
 and for symmetric and antisymmetric actuation, we use \eqref{eq:solvabilityPitchfork} to get 
\begin{equation}
\frac{d^2 A}{dT^2}=b_{1,(\cdot)}\Delta\mu A+b_{2,(\cdot)}\ A^3.
\label{eq:pitchforkFinalForm}
\end{equation}
Equations \eqref{eq:saddleNodeFinalForm} and \eqref{eq:pitchforkFinalForm}  are, respectively, the normal forms for a saddle-node and pitchfork bifurcation \cite{strogatz1994}. They describe the structure of the dynamic equations in the very vicinity of the bifurcation. A simple analysis of these reduced equations gives access to all the dynamical features observed in the vicinity of the transition.

\subsection{Static equilibria}

Equation \eqref{eq:saddleNodeFinalForm} admits two equilibria for $\Delta\mu\leq0$ given by 
\begin{equation}
A_{\textrm{eq}_1}=\displaystyle \sqrt{\frac{a_1}{a_2}}\sqrt{-\Delta \mu},\qquad
A_{\textrm{eq}_2}=\displaystyle -\sqrt{\frac{a_1}{a_2}}\sqrt{-\Delta \mu}.
\label{eq:saddleEquilibriums}
\end{equation}
These  equilibria disappear suddenly at $\Delta\mu=0^+$. They corresponds to the two equilibria S\textsubscript{B} and U\textsubscript{A}, respectively.
The two equilibria in~\eqref{eq:saddleEquilibriums} have the same dependence on $\Delta \mu$,
as observed in Fig. \ref{fig:fig5}D and \ref{fig:fig5}G where we identified only one route to go away from the bifurcation configuration.

Equations \eqref{eq:pitchforkFinalForm} admits three static equilibria
\begin{equation}
A_{\textrm{eq}_1}=0,\quad \forall \ \Delta\mu \in \mathbb{R},\qquad
A_{\textrm{eq}_2}=\displaystyle C_1\sqrt{-\Delta \mu},\quad  
A_{\textrm{eq}_3}=\displaystyle -C_1\sqrt{-\Delta \mu},\quad  \forall \ \Delta\mu \in \left]-\infty,0\right].
\label{eq:pitchforkEquilibriums}
\end{equation}
where $C_1 = \sqrt{b_{1,(\cdot)}/b_{2,(\cdot)}}$, that  correspond to the equilibria U\textsubscript{A}, S\textsubscript{B} and S\textsubscript{A} for the symmetric case and to  the equilibria 
to S\textsubscript{B}, U\textsubscript{B} and U\textsubscript{A} for the antisymmetric case.
Here, $A_{\textrm{eq}_2}$ and $A_{\textrm{eq}_3}$ have the same dependence on $\Delta \mu$, but $A_{\textrm{eq}_1}$ is independent of $\Delta \mu$. 
This is consistent with the observations in Fig. \ref{fig:fig5}E,H (respectively, Fig. \ref{fig:fig5}F,I) where  S\textsubscript{A,B} of the symmetric case (respectively, U\textsubscript{A,B} of the antisymmetric case) go away from the bifurcation following the same route while U\textsubscript{A} of the symmetric case (respectively, S\textsubscript{B} for the antisymmetric case) goes away following a higher order scaling.

\subsection{Linear dynamics}
We perturb around each equilibrium such that  $A(t)=A_{\textrm{eq}_i}+A_\textrm{p}\exp(\sigma t)$, where $A_\textrm{p}$ is the initial (infinitesimal) amplitude of the perturbation and $\sigma$ its growth rate. We substitute back  into \eqref{eq:saddleNodeFinalForm} and linearize to get the standard results
\begin{equation}
\left.
\begin{array}{c}
\sigma_{A_{\textrm{eq}_1}}^2=\displaystyle 2\sqrt{a_1a_2}\sqrt{-\Delta \mu}, \quad\quad
\sigma_{A_{\textrm{eq}_2}}^2=\displaystyle -2\sqrt{a_1a_2}\sqrt{-\Delta \mu}.
\end{array} 
\right.
\label{eq:saddleGrowthRate}
\end{equation}
The first equilibrium $A_{\textrm{eq}_1}$ ($\equiv$ S\textsubscript{B}) possesses two real roots: a negative root that corresponds to a rapidly attenuated mode and a positive that corresponds to a perturbation with an exponentially growing amplitude making $A_{\textrm{eq}_1}$ an unstable equilibrium for \eqref{eq:saddleNodeFinalForm}.
The second equilibrium $A_{\textrm{eq}_2}$ ($\equiv$ U\textsubscript{A}) has two purely imaginary roots implying stable oscillations. 
In Fig. \ref{fig:fig5}A, we compare the eigenvalues in \eqref{eq:saddleGrowthRate} (black lines) to the results obtained from the Cosserat model (green symbols) and Euler beam model (colored lines), showing perfect agreement for $\Delta\mu \ll 1$.

For the symmetric and antisymmetric cases, we find 
\begin{equation}
\left.
\begin{array}{c}
\sigma_{A_{\textrm{eq}_1}}^2=\displaystyle b_{1,(\cdot)}\Delta \mu,\qquad
\sigma_{A_{\textrm{eq}_2}}^2=-2b_{1,(\cdot)}\Delta \mu, \qquad
\sigma_{A_{\textrm{eq}_3}}^2=-2b_{1,(\cdot)}\Delta \mu.
\end{array} 
\right.
\label{eq:pitchforkGrowthRate}
\end{equation}
For the symmetric case, $b_{1,\textrm{sym}}$ is positive, and thus $A_{\textrm{eq}_1}$ (U\textsubscript{A}) admits two purely imaginary roots for $\Delta\mu<0$ that correspond to stable oscillations. For $\Delta\mu\geq 0$, it admits two real roots, one of which is positive, making this equilibrium unstable to the right of the bifurcation.
The two other equilibria $A_{\textrm{eq}_2}$ (S\textsubscript{B}) and $A_{\textrm{eq}_3}$ (S\textsubscript{A}) exist only for $\Delta\mu\leq 0$ and, according to \eqref{eq:pitchforkGrowthRate}, they are both unstable. For the antisymmetric case, $b_{1,\textrm{antisym}}$ is negative, and thus  $A_{\textrm{eq}_1}$(S\textsubscript{B}) is an unstable equilibrium for $\Delta\mu\leq0$. For $\Delta\mu> 0$, this equilibrium admits two purely imaginary roots that correspond to stable oscillations.
The two equilibria $A_{\textrm{eq}_2}$ (U\textsubscript{B}) and $A_{\textrm{eq}_3}$ (U\textsubscript{A}) exist only for $\Delta\mu\leq 0$ and  are stable. 

In Fig. \ref{fig:fig5}B, C, we compare the eigenvalues in \eqref{eq:pitchforkGrowthRate}  (black lines) to the data obtained from the Cosserat (green symbols) and the Euler beam (colored lines) models. For $\Delta\mu\ll 1$, we observe good agreement except for the symmetric case (Fig. \ref{fig:fig5}B) where the numerical data go away from the analytical results, in the very vicinity of the bifurcation. In this region, the system switches from an underdamped to an overdamped regime (see Appendix~\ref{sec:overDampedBoundaryLayer}).

Although the stability analysis carried out in \S\ref{sec:stability} based on the Cosserat and Euler-beam equations provides a better prediction for the linear dynamics of the strip because it is valid even for finite $\Delta\mu$ values, it comes at a cost: it requires to solve a nonlinear eigenvalue problem for each value of $\mu$. This nonlinear system cannot be solved analytically (even for standard Euler buckled configurations $\mu=0$); it is classically solved numerically (see \S\ref{sec:stability} and \cite{nayfeh2008,plaut2009,pandey2014}). 
The eigenvalues \eqref{eq:saddleGrowthRate} and \eqref{eq:pitchforkGrowthRate} obtained from the reduced normal forms provide good analytical estimates of these eigenfrequencies as long as the system is close to the bifurcation.
A similar asymptotic analysis is carried out in \cite{neukirch2012} to predict the vibration frequency of beams beyond the Euler buckling threshold.

\begin{figure}[!t]
 	\centering
 	\includegraphics[width =\textwidth]{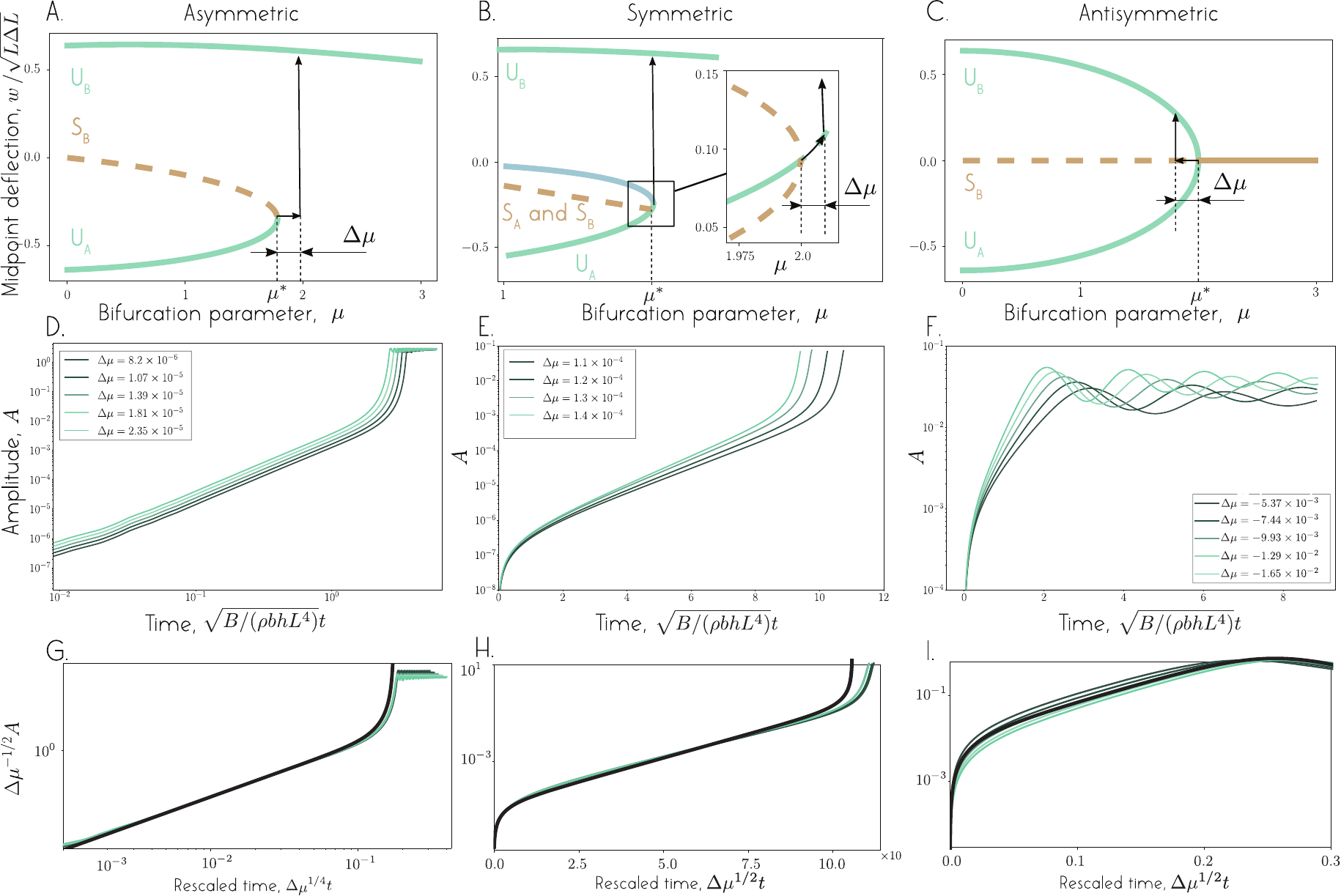}
 	\caption{\footnotesize{\textbf{Snap Through dynamics.} Numerical analysis of the snap-through dynamics associated with (A,D,G)  Asymmetric, (B,E,H) Symmetric, and (C,F,I) Antisymmetric boundary actuation. The snap-through dynamic is analyzed for different values $\Delta\mu$ and represented in term of the evolution of the amplitude $A(t)$. (A-C)  schematic representation of the procedure we followed to move the strip away from the equilibrium at $\mu^\ast$. (D-F) Snap-through dynamic represented on a logarithmic-logarithmic scale (Asymmetric) and on a linear-logarithmic scale (Symmetric and Antisymmetric).  (G-I) Same data represented in term of the rescaled amplitude $\mathcal{A}$ and rescaled time $\tau$.}}
 	\label{fig6}
 \end{figure}

 \section{nonlinear snap-through dynamics}

We analyse the snap-through dynamics obtained from numerical simulations, and compare it to the dynamics described by \eqref{eq:saddleNodeFinalForm} and \eqref{eq:pitchforkFinalForm}. For the asymmetric and symmetric actuation, we study the snap-through dynamics when the system is pulled to the right of the bifurcation ($\Delta \mu >0$), such the equilibrium shape of the strip (U\textsubscript{A}) suddenly disappears or becomes unstable and the strip snaps towards the only remaining stable equilibrium (U\textsubscript{B}). For the antisymmetric actuation, we pull the system to the left of the bifurcation point 
($\Delta \mu <0$) and release it from
the unstable equilibrium S\textsubscript{B}, causing the strip to snap towards the stable equilibrium U\textsubscript{A} (Fig. \ref{fig6}A-C).

\subsection{Asymmetric actuation} 

In our Cosserat simulations, we obtain the snapping dynamics for different values of $\Delta\mu$ by employing a technique similar to that introduced experimentally in \cite{gomez2017}. We start with the strip equilibrium configuration $(W_\textrm{eq}^\ast, \Lambda_\textrm{eq}^\ast)$ at the bifurcation point $\mu^{\ast}$. We then pull the system to the right of the bifurcation by increasing the angle applied at the left boundary to the value $\mu=\mu^{\ast}+\Delta\mu$. During this process, the midpoint of the strip is maintained at its initial position by applying an additional boundary condition at the centerline of the Cosserat rod (which plays the role of the indenter used in \cite{gomez2017}). When we reach the desired value of $\mu$, the midpoint constraint is suddenly released (after waiting enough time for the strip to reach equilibrium), and the strip snaps to the U\textsubscript{B} configuration.  

From these simulations, we {contrust} the  evolution of $A(T)$. The process is repeated for several values of $\Delta \mu$ (Fig. \ref{fig6}D). In Fig. \ref{fig6}G, as done in \cite{gomez2017}, we show the results in non-dimensional form  by plotting $\mathcal{A}=\Delta \mu^{-1/2}A$ as a function of  $\tau=\Delta \mu^{1/4}T$ which are the natural spatial and temporal scales in the vicinity of the bifurcation. Clearly, all the data collapse on the same master curve. 
We compare these data to the dynamics described by \eqref{eq:saddleNodeFinalForm} (black lines). For this purpose, we integrate \eqref{eq:saddleNodeFinalForm} in time using a 4th order Runge-Kutta (RK4) integrator with initial conditions $(A(T=0),dA(T=0)/dT=0)$. The dynamics of the normal forms compares well with the Cosserat data (green lines) at short time. At larger time however, as explained in \cite{gomez2017}, the dynamics described by \eqref{eq:saddleNodeFinalForm} blows off to infinity while the numerical data plateau. This plateau is observed when the strip reaches the new equilibrium U\textsubscript{B}. The latter is far from the bifurcation point (Fig. \ref{fig6}A) and is therefore not captured by the asymptotic analysis (U\textsubscript{B} does not appear on the bifurcation diagram associated with the reduced form in \cite{radisson2022PRL}). The saturation observed in the numerics comes from the role played by higher order terms that are neglected in the asymptotic analysis but that become dominant as soon as the conditions $\Delta\mu\ll 1$, $\Delta W\ll1$ and $
\Delta \Lambda\ll1$ are no longer satisfied. 

\subsection{Symmetric actuation}

 We conduct a similar analysis: for each value of $\Delta \mu$, we start from the unstable configuration U\textsubscript{A} and analyse how the strips snaps towards the stable configuration U\textsubscript{B}. We carry out Cosserat simulations where we initialize the strip using the solution obtained from the static Euler-beam equation, which does not satisfy the discrete Cosserat equations leading to transient numerical shocks. After these spurious initial shocks, we observe the snapping dynamics of the strip (\ref{fig6}E). In Fig. \ref{fig6}H, we plot the obtained evolution of $\mathcal{A}=\Delta\mu^{-1/2}A$ as a function of $\tau=\Delta\mu^{1/2}T$. These are the natural spatial and temporal time scales in the vicinity of a pitchfork bifurcation. The data shown in the figure are taken after the spurious initial shocks have disappeared. These numerical data are compared to the dynamics described by  \eqref{eq:pitchforkFinalForm} (black line) with initial conditions $(A(T=0)=0, dA(T=0)/dT=v_0)$, where $v_0$ is the initial speed obtained from the numerical simulations after the initial shocks have disappeared. This initial velocity is small but non-zero and is responsible for the initial kick observed in Fig. \ref{fig6}E, H. At early time, the dynamics is linear and the amplitude grows as the sum of the two exponential modes given in \eqref{eq:pitchforkGrowthRate}. After this initial phase, the amplitude blows off to infinity due to the destabilizing cubic term in \eqref{eq:pitchforkFinalForm}. The numerical data, however, plateau when the strip reaches the new equilibrium U\textsubscript{B}. As for the asymmetric case, the latter is far from the bifurcation point and is not captured by our asymptotic analysis.

\subsection{Antisymmetric actuation}
 We study the dynamics of the strip starting from the unstable equilibrium S\textsubscript{B} to the left of the bifurcation. To initialize the simulations, we follow the same procedure as the one employed for the symmetric case. The numerical data are shown in Fig. \ref{fig6}F for different $\Delta\mu$ values. In Fig. \ref{fig6}I, these data are rescaled and plotted as $\mathcal{A}=\Delta\mu^{-1/2}A$ in term of $\tau=\Delta\mu^{1/2}T$ and compared to the dynamics described by \eqref{eq:pitchforkFinalForm} (black line). 
The early dynamic of snapping follows a similar pattern as the one obtained for the symmetric case. At early time, the dynamic is linear and $A(T)$ grows as the sum of the two independent modes given in \eqref{eq:pitchforkGrowthRate}. Then, $A(T)$ reaches a plateau and oscillates when the cubic term in \eqref{eq:pitchforkFinalForm} saturates the linear term (Fig. \ref{fig6}F,I). Contrary to the two other cases, the saturation observed when the strip reaches the new equilibrium (U\textsubscript{A} or U\textsubscript{B}) is well captured here. This is because the first and only nonlinear term considered in our asymptotic analysis is stabilizing, whereas it is destabilizing in the asymmetric and symmetric cases where the saturation comes from higher order nonlinear terms.

\section{Discussion}

The analysis presented in this paper relates the dynamic characteristics of an elastic structure in the vicinity of a shape transition to the nature of the underlying bifurcation. In particular, the critical slowing down in the vicinity of the bifurcation $\mu^\ast$ is responsible for the extreme sensitivity of the snap-through time to the external control parameter $\Delta \mu = \mu - \mu^\ast$ \cite{gomez2017} and follows a precise scaling that depends on the type of bifurcation. This scaling is a universal feature of the corresponding bifurcation and is observed near such bifurcations in different contexts, including turbulence intermittency \cite{pomeau1980,berge1986} and electronics \cite{sone1985,trickey1998}. 

To demonstrate the utility of our analysis, we discuss how the critical slowing down properties could be exploited to anticipate when the system is approaching a bifurcation and even to predict the exact position of the bifurcation point $\mu^\ast$ when it is not known a priori.

According to our analysis, the typical time scale $T=1/\sqrt{|\sigma^2|}$ associated with the dynamics of how the strip diverges  away from its equilibrium shape near the bifurcation $\mu^\ast$ scales as  $T \propto \Delta\mu^{-a}$, equivalently, the distance from the bifurcation scales as $\Delta\mu \propto T^{-1/a}$.
Theoretically, this knowledge, together with knowing the value of $a$, can be used to predict the location of the bifurcation $\mu^\ast$: by performing multiple simulations at various values of $\mu$ (say using the analysis in Fig.~\ref{fig:fig4} and~\S\ref{sec:stability}), one can calculate $\sigma^2$ and $T =1/\sqrt{|\sigma^2|}$ and plot $\Delta\mu = T^{-1/a}$ as a function of $\mu$. By extrapolation, the intercept where $\Delta \mu =T^{-1/a} =0$ gives the value of the bifurcation $\mu^\ast$.
However, $T(\mu)$ follows the expected scaling only in the vicinity of the bifurcation (see Fig. \ref{fig:fig5}). Thus, to predict the position of the shape transition accurately using this extrapolation method, one must make sure that the system is close enough from the bifurcation point and that $T(\mu)$ follows the expected scaling. This is challenging given only the value of the bifurcation parameter $\mu$, without knowing the distance to the bifurcation $\Delta \mu$.

A remedy is readily available: the critical slowing down is known to be a robust early warning signal of the vicinity of a transition \cite{scheffer2009, dakos2008, scheffer2012, dai2012}. Thus, by measuring $T(\mu)$ at different values of $\mu$, one should notice a significant increase of the characteristic time scale when approaching the bifurcation. This feature can be exploited to determine if the system is already close enough to the bifurcation and to predict the position of the bifurcation by extrapolating $T^{-1/a}(\mu)$. We next apply this approach to the asymmetric, symmetric, and antisymmetric systems for which we have already determined the position $\mu^*$ of the bifurcation to probe the efficacy of such algorithm in anticipating the value of $\mu^*$.

\begin{figure}[!t]
	\centering
	\includegraphics[width=\linewidth]{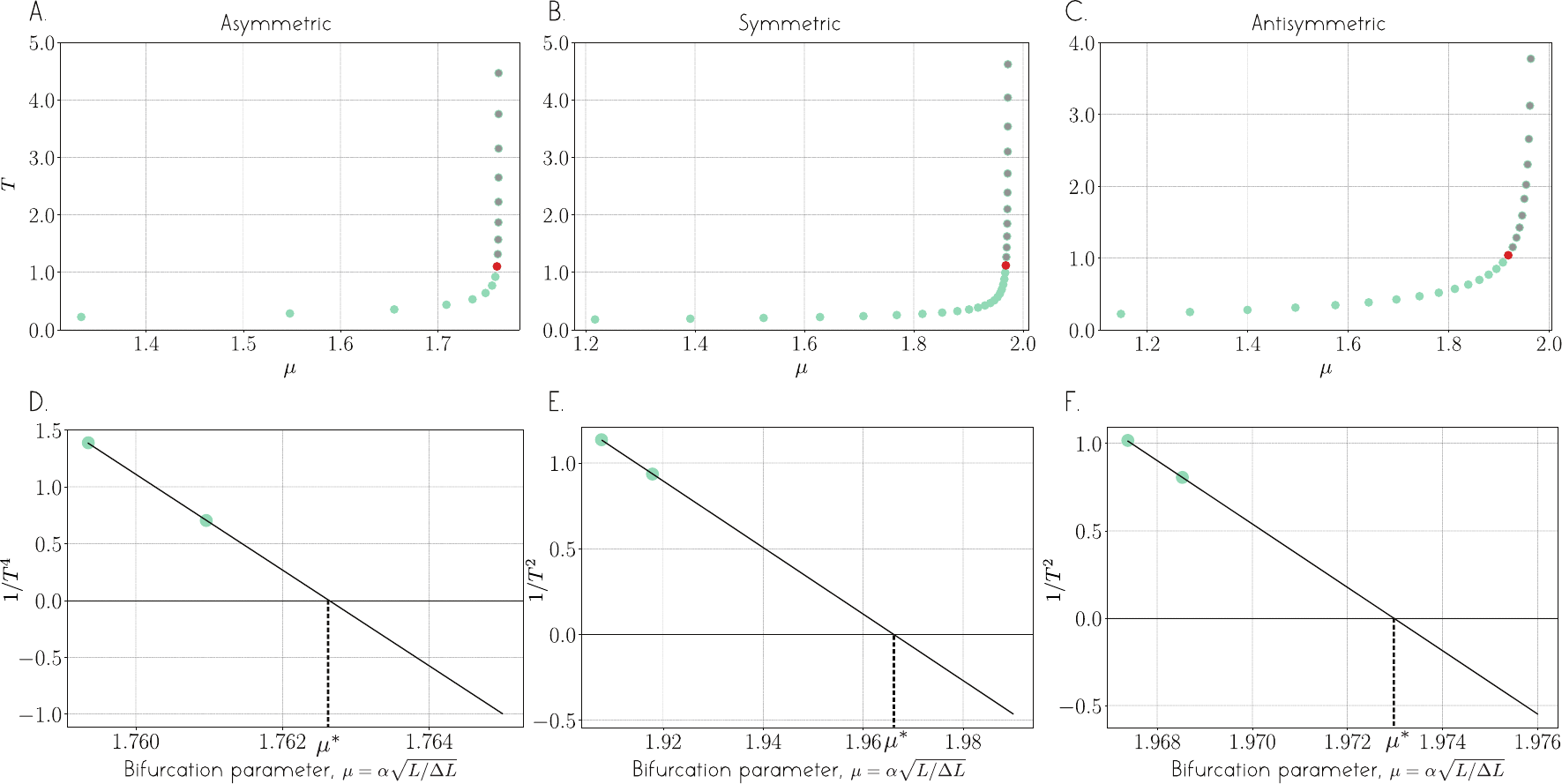}
	\caption{\footnotesize{\textbf{Anticipate shape transitions} (A, B, C) We measure the typical time scale associated with the dynamic of the strip (i.e period of the fundamental mode of vibration around a stable equilibrium) using Cosserat simulations. We note a significant increase of this time scale when a certain value of $\mu$ is approached (green dots). This indicate that the system is getting close to a shape transition. When the time scale becomes about ten times larger than the time scale far from the bifurcation (red dot), we use this measurement and the previous one (D, E, F) to evaluate the function $f(\mu)=1/\tau^{1/a}$ and extrapolate linearly until $f(\mu)=0$. This provides an estimation of $\mu^*$ the bifurcation point. The grey dot symbols are known from the previous study but are not exploited here for this prediction.}}
	\label{fig:anticipatingShapeTransitions}
\end{figure}

\begin{table}[!t]
\caption{Comparison of the predicted value $\mu^{*}_{\text{extrap}}$ for the bifurcation point with the actual value $\mu^*$ obtained from our numerical simulations.}
\begin{tabular}{c| ccc}
& \textbf{Asymmetric} & \textbf{Symmetric} & \textbf{Antisymmetric}\\[3mm]
\toprule
$\mu^*$&$1.762708132$&$1.972550$&$1.9670$\\[3mm]
$\mu^{*}_{\text{prediction}}$&$1.762637064$&$1.972976$&$1.9661$\\[3mm]
\hline
\end{tabular} 
\label{tab:mubPrediction}
\end{table}

We start from $\mu=0$, we measure the typical time scale $T(\mu)$ associated with the fundamental mode of vibration (Fig.~\ref{fig:fig3} and  \S\ref{sec:stability}) for increasing values of $\mu$. In Fig. \ref{fig:anticipatingShapeTransitions} A-C, we plot, for each actuation, the typical time scale of the system obtained from the numerical measurements performed in\S\ref{sec:stability}. We assume that we do not have access to the data points in grey and stop the measurements when $T(\mu)$ starts to significantly increase.  As a criteria, we increase $\mu$ until $T(\mu)$ exceeds its value far from the bifurcation (here, the value at $\mu=0$) by one order of magnitude. The first measurements that satisfies this criteria is highlighted in red in Fig. \ref{fig:anticipatingShapeTransitions}. We use this last measurement and nearest measurement to the left to calculate $T^{-1/a}(\mu)$ and linearly extrapolate until we find the zero-intercept (Fig. \ref{fig:anticipatingShapeTransitions} D-F). The  value of $\mu$ for which  $T^{-1/a}(\mu)=0$ gives an estimate for the bifurcation point. In Table \ref{tab:mubPrediction}, we compare the values $\mu^{*}_{\text{predication}}$ predicted from this method to the actual value $\mu^*$ reported in Table~\ref{tab:muStarNumRot}.
 These estimates can be refined further by using more measurements closer to the bifurcation point. We note that a similar method was proposed in \cite{gomez2018c} to estimate the pull-in voltage in MEMS devices. However, the estimation there was performed using the snap-through time to the right of the bifurcation. Our method instead, can anticipate the transition without taking the system through the bifurcation point.
 
 This analysis depends on apriori knowledge of the scaling. 
 The value of $a$ depends only on the type of bifurcation the system undergoes, which can be predicted from symmetry-breaking considerations~\cite{radisson2022}, with $a=1/4$ for a saddle-node and $a=1/2$ for a pitchfork. 
This scaling is valid for underdamped systems (for an overdamped system, $a=1/2$ for a saddle-node and $a=1$ for a pitchfork; see Appendix \ref{sec:overDampedBoundaryLayer}).
Thus, to employ this kind of predictive analysis, in addition to knowing the type of transition, one has to carefully check whether the system is underdamped or overdamped.

\bibliography{referencesPRE}
	
	
\appendix
\renewcommand\thefigure{\thesection.\arabic{figure}}  
\setcounter{figure}{0}
\renewcommand\thetable{\thesection.\Roman{table}}  


\section{Snap-through time}
\label{sec:snappingTime}

For asymmetric actuation, Gomez \textit{et al.} measured the typical snap-through time, the time to transition from U\textsubscript{A} to U\textsubscript{B}, when the system is pulled by a distance $\Delta\mu$ to the right of the bifurcation (see supplemental document of \cite{gomez2017}). In the experiments of~\cite{gomez2017}, the strip was carefully placed at the bifurcation and the mid-point of the strip was maintained at its bifurcation position using an indenter while $\mu$ was varied by pulling the system to the right of the bifurcation, where no equilibrium is available, by a distance $\Delta\mu$. The indenter was suddenly released letting the strip free to snap towards the far away equilibrium U\textsubscript{B}. In Fig.~\ref{fig:fig4}, we obtained the slowing down scaling by analyzing the linear dynamics of the strip around the equilibria to the left of the bifurcation point. Here, we show that the same kind of procedure as the one introduced experimentally by Gomez \textit{et al.} can be exploited in our numerical simulations.

\begin{figure}[!t]
	\centering
	\includegraphics[width=0.8\linewidth]{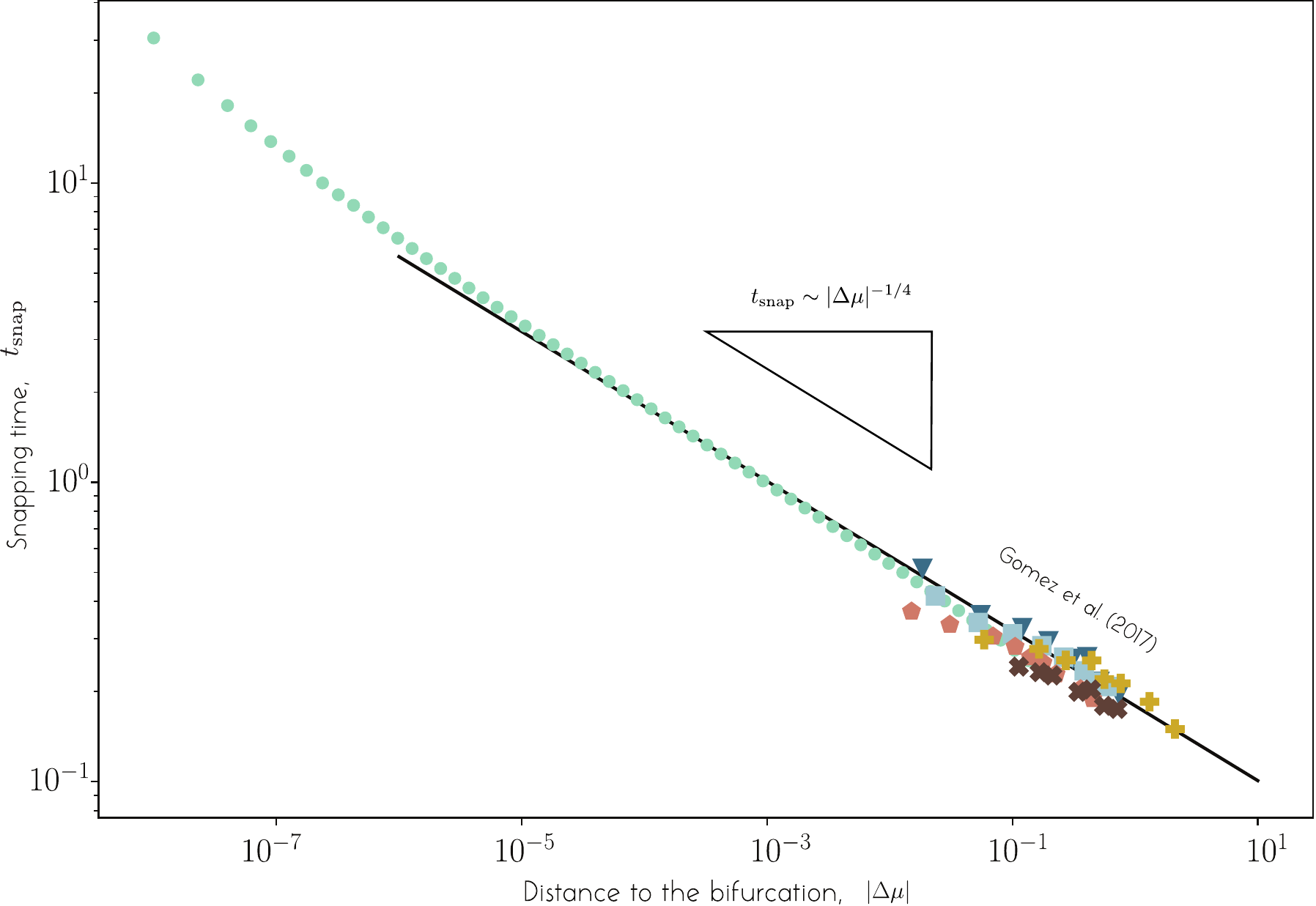}
	\caption{\footnotesize{\textbf{Snapping time.} Comparison of snapping times when the system is placed to the right of the bifurcation by a distance $\Delta\mu$ between our 3D Cosserat simulations and the experiments carried out by Gomez et al. \cite{gomez2017}. Their experimental data were obtained with strips made of PET with $L=240mm$  $\alpha=21.34^{\circ}$ (PET1), $L=290mm\ \alpha=19.85^{\circ}$ (PET2) and $L=430mm\ \alpha=21.17^{\circ}$ (PET3) and strips made from steel with $L=280mm\ \alpha=17.14^{\circ}$ (Steel1) and $L=140mm$  $\alpha=22.51^{\circ}$ (Steel2). The black line represents the analytical prediction they obtained.}}
	\label{fig:fig7}
\end{figure}

We use the static equilibrium at the bifurcation point $\mu^*$ as initial condition, and maintain the mid-point vertex at this position by constraining the center line of the Cosserat rod to remain at this initial position while the angle $\mu$ at the left endpoint is increased. When the target value of $\mu$ is reached, the mid-point constraint is suddenly removed letting the strip free to snap to the U\textsubscript{B} configuration. The snapping time is taken as the duration between the time at which the constraint is released and the time at which the mid-point position first hits its final equilibrium position. The resulting non-dimensional snapping time ($t_{\text{snap}}$) is reported in Fig. \ref{fig:fig7} (green dots) in term of the distance $\Delta\mu$ to the bifurcation and compared to the experimental data obtained by Gomez \textit{et al.}~\cite{gomez2017}.  Our numerical data collapse almost perfectly on their analytic prediction (black line) for the snapping time (i.e., the time for which the amplitude of the leading order mode diverges to infinity) except very close to the  bifurcation. In this region, the system passes from an underdamped to an overdamped regime (see Appendix \ref{sec:overDampedBoundaryLayer}). 

These data validate the numerical method exploited here and in the companion paper~\cite{radisson2022} to solve the nonlinear Cosserat equations. They also highlight an advantage for carrying out numerical simulations as opposed to real experiments: 
numerical simulations allow to analyze the dynamics of the strip much closer to the bifurcation ($\Delta\mu \ll 1$) than accessible experimentally.


\section{Over-damped boundary layer} \label{sec:overDampedBoundaryLayer}

The numerical data (green symbols) presented in Fig. \ref{fig:fig5}B and Fig.\ref{fig:fig7} seem to deviate from the analytical prediction (black lines) obtained from the reduced equations in~\S\ref{sec:asymptoticAnalysis} and~\S\ref{sec:Aeq}. This difference in behavior is associated with dissipation mechanisms, that despite being small, become predominant in the very vicinity of the bifurcation due to the critical slowing down. 

In our numerical simulations, a small damping term is added to mimic material dissipation in the elastic strip. Taking this effect into account in our analysis would introduce a damping term into \eqref{eq:beam_equation_nodim} such that
\begin{equation}
\frac{\partial^2 W}{\partial T^2}+\xi\frac{\partial W}{\partial T}+\frac{\partial^4 W}{\partial X ^4}+\Lambda^2 \frac{\partial^2 W}{\partial X^2}=0.
\label{eq:beam_equation_damping}
\end{equation}
Here, $\xi=\nu L^2/\sqrt{\rho b h B}$ with $\nu$ being the dynamic viscosity coefficient used in our numerical implementation of the Cosserat equations, is a non dimensional parameter that compares viscous forces acting over the inertial time scale $\sqrt{\rho b h L^4/B}$ to inertial forces \cite{gomez2018}. For large $\xi$, the second term in the LHS of \eqref{eq:beam_equation_damping} dominates the dynamic. For small $\xi$,  this term is small and can be neglected. In all the numerical experiments carried out in this paper and in the experiments carried out in \cite{gomez2017}, $\xi$ is small and \eqref{eq:beam_equation_nodim} can be used instead of \eqref{eq:beam_equation_damping}.
In \cite{gomez2018}, the authors demonstrated that this is not true in the very vicinity of the bifurcation. Close to the bifurcation, the dynamics slows down and even for $\xi\ll 1$ there is a boundary layer of thickness $\Delta\mu\sim\xi^{1/a}$ where the viscous term is not negligible and where the dynamic becomes over-damped \cite{gomez2018}. Outside this boundary layer (i.e $\Delta\mu\gg\xi^{1/a}$), the second term in \eqref{eq:beam_equation_damping} is negligible and the dynamics is governed by \eqref{eq:beam_equation_nodim}; the analysis carried out in the main paper is valid.

In the boundary layer (i.e $\Delta\mu\ll\xi^{1/a}$), the first term in \eqref{eq:beam_equation_damping} can be neglected and the dynamics is governed by (here $\Bar{T}=\xi^{-1}T$)
\begin{equation}
\frac{\partial W}{\partial \Bar{T}}+\frac{\partial^4 W}{\partial X ^4}+\Lambda^2 \frac{\partial^2 W}{\partial X^2}=0.
\label{eq:beam_eq_overDamped}
\end{equation}
We repeat the stability analysis done in the main text in this overdamped case. We obtain the exact same eigenvalue problem but with eigenvalue $\sigma$ instead of $\sigma^2$. This yields values for $a$ that are twice larger than the ones found in the main text (i.e we get $a=1/2$ for the asymmetric case and $a=1$ for the two other cases). We note that the scaling observed for our numerical data in Fig. \ref{fig:fig5}B very close to the bifurcation corresponds to $a=1$ indeed.

We then perform the asymptotic analysis, starting with the first order in time beam equation and these new values obtained for $a$. We obtain the first order in time version of \eqref{eq:saddleNodeFinalForm} and \eqref{eq:pitchforkFinalForm}. We note that the values $a=1/2$ and $a=1$ obtained from the stability analysis correspond to the well known exponent that characterises the critical slowing down near first order in time saddle-node and pitchfork, respectively. 

In our numerical simulations in Fig. \ref{fig:fig5}B and Fig. \ref{fig:fig7} the value of $\xi$ is $\xi_1\approx 2\times 10^{-4}$ and $\xi_2\approx 6\times 10^{-4}$, respectively. From these values and prior knowledge of the value of $a$  we can estimate the thickness of the boundary layer $\delta=\xi^{1/a}$. With $a=1$ and $a=1/2$ (first order pitchfork and first order saddle node, respectively) this yields $\delta_1=\xi_1^{1/a}\approx 2\times 10^{-4}$ for the parameters used in Fig. \ref{fig:fig5}B and $\delta_2=\xi_2^{1/a}\approx 3.2\times 10^{-7}$ for those used in Fig. \ref{fig:fig7}. We note that these values correspond quantitatively to the minimum value $\Delta\mu$ under which the numerical data starts to go away from the under-damped theory (black lines).

\begin{figure}[!t]
 	\centering
 	\includegraphics[width =\textwidth]{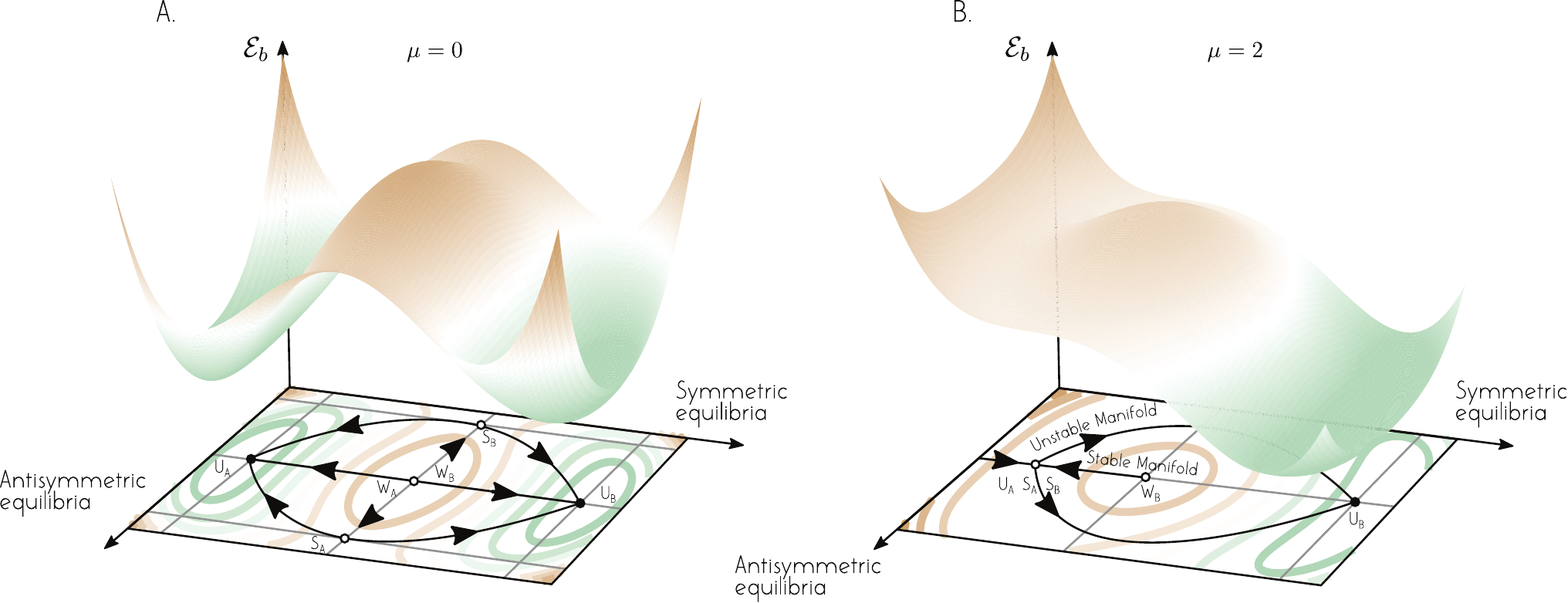}
 	\caption{\footnotesize{\textbf{Energy landscape} The bending energy of the three first pairs of equilibrium states (U\textsubscript{A}, U\textsubscript{B}, S\textsubscript{A}, S\textsubscript{B}, W\textsubscript{A}, W\textsubscript{B}) is computed and plotted on a 2D space spanned by $W^S_{0}$ and $W^A_{3/4}$ (see main text for definition). The energy surface between the equilibria is drawn arbitrarily. \textbf{A.} At $\mu=0$, the energy landscape is typical of the bi-stable Euler buckled system, with two potential wells U\textsubscript{A} and U\textsubscript{B} separated by two lowest energy barriers S\textsubscript{A} and S\textsubscript{B}. The bump at the origin corresponding to the two higher energy barriers W\textsubscript{A} and W\textsubscript{B}. \textbf{B.} At $\mu=2$, under the symmetric actuation, S\textsubscript{A} and S\textsubscript{B} become symmetric and merge with U\textsubscript{A}. The latter becomes unstable relatively to an antisymmetric perturbation. However, it is still stable relatively to a symmetric mode of perturbation. The resulting equilibrium is therefore a saddle with a (actually, an infinite number of) stable and an unstable manifold.}}
 	\label{fig:fig6}
 \end{figure}

\section{Pre snap-through oscillations}
\label{app:sym_osc}
In Fig.~\ref{fig:fig2}H and I, 
the exponential snapping dynamics predicted in \cite{gomez2017} and described by \eqref{eq:pitchforkFinalForm} is preceded by damped oscillations (see Fig. \ref{fig:fig2}H and \ref{fig:fig2}I). 
To elucidate the origin of these pre-snap-through oscillations, we plot in Fig.~\ref{fig:fig6} a simplified energy diagram of the symmetrically actuated strip at $\mu=0$ (both boundaries are at a zero angle with the horizontal) and at $\mu=2$ (first bifurcation).
Specifically, we compute the bending energy  $\mathcal{E}_b=EI/2\int_{-1/2}^{1/2} (\partial^2w/\partial x^2)^2dx$ at each of the six first static equilibrium modes (U\textsubscript{A}, U\textsubscript{B}, S\textsubscript{A}, S\textsubscript{B}, W\textsubscript{A}, W\textsubscript{B}) based on the static analysis in section~\ref{sec:EB}. We then represented these energy values as a function of a 2D space, where one direction spans the mid-point deflection of the symmetric modes (U, W)  and the second direction spans the  deflection at $X=3/4$ of the antisymmetric modes (S) of buckling (see Fig. \ref{fig:fig6}). The antisymmetric component of the symmetric modes are set to zero and vice versa. 

At $\mu=0$, the energy diagram displays two potential wells that correspond to the two buckled equilibria U\textsubscript{A} and U\textsubscript{B}, which are symmetric under the transformation $w \rightarrow -w$ and have the same bending energy. Meanwhile, S\textsubscript{A} and S\textsubscript{B} are symmetric under the transformation $x \rightarrow -x$ and occupy a higher level of bending energy. Lastly, W\textsubscript{A} and W\textsubscript{B} are symmetric under the transformation $w \rightarrow -w$ with zero mid-point deflection and occupy an even higher energy level.  When the boundaries of the strip are rotated, this standard energy landscape is reshaped until one (or both) of these two lowest energy barriers "breaks" therefore allowing the system to transition from one state to another.

In the case of symmetric boundary actuation, the energy levels of U\textsubscript{A} and W\textsubscript{A} increase, while those of U\textsubscript{B} and W\textsubscript{B} decrease and those of S\textsubscript{A} and S\textsubscript{B} remain the same. At $\mu=2$,
the three equilibria U\textsubscript{A}, S\textsubscript{A} and S\textsubscript{B} merge together at the same energy level  in a subcritical pitchfork bifurcation (Fig. \ref{fig:fig2}). The fundamental mode of perturbation of U\textsubscript{A} becomes unstable (Fig.~\ref{fig:fig2}H and Fig. \ref{fig:fig4}B where the first eigenvalue of U\textsubscript{A} crosses the zero axis). 
This eigenvalue is associated with an antisymmetric mode of perturbation $W_p(X)$ (\S\ref{sec:stability}) and therefore at this point two routes (depending on the sign of the amplitude $\epsilon$ of this antisymmetric mode) are available for the strip to snap from U\textsubscript{A}  to U\textsubscript{B} (Fig. \ref{fig:fig6}). However, the first harmonic of U\textsubscript{A}, labelled U\textsubscript{A1} in Fig. \ref{fig:fig4}B, still has a negative eigenvalue and is stable. This second eigenvalue is associated with a symmetric $W_p(X)$. 
This means that the unstable equilibrium born from the merging of U\textsubscript{A}, S\textsubscript{A} and S\textsubscript{B} is a saddle and possesses (actually an infinity of) a stable and an unstable manifold (Fig. \ref{fig:fig6}B). At the first bifurcation, when $\mu$ is suddenly increased from a value slightly smaller to a value slightly higher than $\mu^*$ the route towards U\textsubscript{B} suddenly opens (a stable manifold (antisymmetric) is turned into an unstable one). But at the same time the "kick" imposed by the actuation is symmetric and pushes the strip along the stable manifold (symmetric) where the route towards U\textsubscript{B} is 'closed'. As the system is placed slightly to the right of the bifurcation, the first eigenvalue becomes positive (unstable) but is infinitesimal (due to the critical slowing down at the bifurcation) while the second eigenvalue is still negative (stable) and has a finite value (inset Fig. \ref{fig:fig4}B). Therefore, the system has time to oscillate along the stable manifold before being attracted along the unstable one. Increasing $\mu$ a little more, this stable manifold becomes unstable at a secondary bifurcation. The eigenvalue corresponding to the first harmonic of perturbation of U\textsubscript{A} vanishes with the eigenvalue associated with the fundamental mode of perturbation of W\textsubscript{B} at $\mu\approx2.012$ where these two equilibria suddenly disappear in what resembles a saddle-node bifurcation. This is confirmed by looking at how the absolute values of these two eigenvalues decrease when approaching this secondary bifurcation (Fig. \ref{fig:fig5}B). Note that the slowing down at this secondary bifurcation follows the saddle-node scaling  (i.e $\sqrt{|\sigma^2|}\sim |\Delta\mu|^{1/4}$). 

These pre-snapping oscillations are often observed in step-loaded arches (e.g \cite{das2009, chen2014}). They are usually described as ``indirect snap-through'', a mechanism where a mode of oscillation acts as a parametric forcing to another mode and triggers a parametric resonance that leads to snap-through \cite{lock1966}. This is not the case here, where the oscillations that precede snap-through are fully described by purely linear mechanisms.

\section{Asymptotic analysis for asymmetric actuation}\label{sec:asymptotic_asym}

We recall the asymptotic analysis of Gomez et al. \cite{gomez2017} for the asymmetric boundary actuation, where an order reduction method is used to reduce the infinite degree of freedom system  to a single degree of freedom system: the amplitude of the leading order mode in the vicinity of the bifurcation. 

Substituting $a = 1/4$ and $b=c=1/2$ into~\eqref{eq:expansion} leads to
\begin{equation}
\begin{split}		
W(X,\tau)&=W_{\textrm{eq}}^*(X)+\Delta\mu^{1/2} W_{0}(X,\tau)+\Delta\mu W_{1}(X,\tau)+O(\Delta \mu^{3/2}),\\
\Lambda(\tau)&=\Lambda_{\textrm{eq}}^*+\Delta\mu^{1/2}\Lambda_{0}(\tau)+\Delta\mu\Lambda_{1}(\tau)+O(\Delta \mu^{3/2}). 
\end{split} 
\label{eq:expansionSN}
\end{equation}
In turn, substituting \eqref{eq:expansionSN} into \eqref{eq:beam_slow_bifurcation} and arranging terms order-by-order leads to the following analysis.

\bigskip
\par\noindent
\textbf{At Order $\Delta \mu^{1/2}$.}
At the leading order, the system is of the form
\begin{equation}
\begin{split}
\displaystyle \frac{\partial^4 W_0}{\partial X^4}+(\Lambda_{\textrm{eq}}^*)^2\frac{\partial^2 W_0}{\partial X^2}+2\Lambda_{\textrm{eq}}^*\Lambda_0\frac{d^2 W_{\textrm{eq}}^*}{d X^2}=0,
\qquad
\displaystyle \int_{0}^{1}\frac{d W_{\textrm{eq}}^*}{d X}\frac{\partial W_0}{\partial x}dX=0, 
\\[3mm]
W_0(0)=W_0(1)=0, 
\qquad \displaystyle\left.\frac{\partial W_0}{\partial X}\right|_{X=0}=0, 
\qquad \displaystyle\left.\frac{\partial W_0}{\partial x}\right|_{X=1}=0.
\end{split} 
\label{eq:leadingOrderSystem}
\end{equation}
Following Gomez et al., 
we define the linear operator $\mathcal{L}(W_i, \Lambda_i)$,
\begin{equation}
\mathcal{L}(W_i,\Lambda_i)\equiv\frac{\partial^4 W_i}{\partial X^4}+(\Lambda_{\textrm{eq}}^*)^2\frac{\partial^2 W_i}{\partial X^2}+2\Lambda_{\textrm{eq}}^*\Lambda_i\frac{d^2 W_{\textrm{eq}}^*}{d X^2},
\label{eq:theOperator}
\end{equation}
and we write the leading order term in the asymptotic expansion $(W_0(X,\tau), \Lambda_{0}(\tau))$ as in~\eqref{eq:leadingmode}; that is, we write 
   $ W_0(X,\tau) = \mathcal{A}(\tau)\Phi_0(X)$, and  $\Lambda_{0}(\tau)=\mathcal{A}(\tau)$,
where $\Phi_0(X)$ is the shape of the leading order mode and $\mathcal{A}(\tau)$ its amplitude.
Substituting~\eqref{eq:leadingmode} into\eqref{eq:leadingOrderSystem} and noting that $\Lambda_0$ is independent of $X$, we get
\begin{equation}
\begin{split}
\displaystyle \mathcal{A}(\tau)\left(\frac{d^4 \Phi_0}{d X^4}+(\Lambda_{\textrm{eq}}^*)^2\frac{d^2 \Phi_0}{d X^2}+2\Lambda_{\textrm{eq}}^*\frac{d^2W_{\textrm{eq}}^*}{d X^2}\right)=0,
\qquad
\displaystyle \mathcal{A}(\tau)\int_{0}^{1}\frac{d W_{\textrm{eq}}^*}{d x}\frac{d \Phi_0}{dX}dX=0, \\[3mm]
\Phi_0(0)=\Phi_0(1)=0, 
\qquad 
\displaystyle\left.\frac{d \Phi_0}{dX}\right|_{X=0}=0,
\qquad
\displaystyle\left.\frac{d \Phi_0}{d X}\right|_{X=1}=0.
\end{split} 
\label{eq:leadingOrderSystem_separationOfvariable}
\end{equation}
The solution of the ODE in \eqref{eq:leadingOrderSystem_separationOfvariable} is given by
\begin{equation}
\Phi_0(X)=A_0\sin(\Lambda_{\textrm{eq}}^* X)+B_0\cos(\Lambda_{\textrm{eq}}^* X)+C_0 X+D_0+\frac{1}{\Lambda_{\textrm{eq}}^*}X\frac{dW_{\textrm{eq}}^*}{dX}.
\label{eq:leadingOrderSystem_sol}
\end{equation}
Making use of the boundary conditions at $X=0$ for $\Phi_0$, $d\Phi_0/dX$, and $dW_{\textrm{eq}}^*/dX$, we get
\begin{equation}
\Phi_0(X)=\frac{1}{\Lambda_{\textrm{eq}}^*}(X\frac{dW_{\textrm{eq}}^*}{dX}-\mu^* X)+A_0\left(\sin(\Lambda_{\textrm{eq}}^* X)-\Lambda_{\textrm{eq}}^* X\right)+B_0\left(\cos(\Lambda_{\textrm{eq}}^* X)-1\right).
\label{eq:leadingOrderSystem_solUsingleftBcs}
\end{equation} 
This is the leading order solution obtained by Gomez et al. (see \cite{gomez2017} Supplemental Material).
The expressions of the constants $A_0$ and $B_0$ are then obtained from the boundary conditions at the right end of the strip
\begin{equation}
\begin{array}{c}
\displaystyle \left. W_{\textrm{eq}}^*\right|_{X=0}= \left. W_{\textrm{eq}}^*\right|_{X=1}=\left.\frac{dW_{\textrm{eq}}^*}{dX}\right|_{X=1}=0 \qquad \displaystyle\left.\frac{d W_{\textrm{eq}}^*}{dX}\right|_{X=0}=\mu^*,
\end{array}
\end{equation}
which yields
\begin{equation}
\left.
\begin{split}
A_0&=\displaystyle \frac{-2\mu^* \sin^2(\Lambda_{\textrm{eq}}^*/2)\left\{\left((\Lambda_{\textrm{eq}}^*)^2-2\right)\cos(\Lambda_{\textrm{eq}}^*)-2\Lambda_{\textrm{eq}}^*\sin(\Lambda_{\textrm{eq}}^*)+2\right\}}{(\Lambda_{\textrm{eq}}^*)^2\left\{2\cos(\Lambda_{\textrm{eq}}^*)+\Lambda_{\textrm{eq}}^*\sin(\Lambda_{\textrm{eq}}^*)-2\right\}^2},\\[5mm]
B_0&=\displaystyle \frac{-\mu^*\left\{(\Lambda_{\textrm{eq}}^*)^3+(\Lambda_{\textrm{eq}}^*)^2\sin(\Lambda_{\textrm{eq}}^*)\left[\cos(\Lambda_{\textrm{eq}}^*)-2\right]+2\left[\Lambda_{\textrm{eq}}^*\cos(\Lambda_{\textrm{eq}}^*)-\sin(\Lambda_{\textrm{eq}}^*)\right]\left[\cos(\Lambda_{\textrm{eq}}^*)-1\right]\right\}}{(\Lambda_{\textrm{eq}}^*)^2\left[2\cos(\Lambda_{\textrm{eq}}^*)+\Lambda_{\textrm{eq}}^*\sin(\Lambda_{\textrm{eq}}^*)-2\right]^2}.
\end{split}
\right.
\end{equation}
Equation \eqref{eq:leadingOrderSystem_solUsingleftBcs} describes the shape of the leading order mode in the very vicinity of the bifurcation (i.e the route the strip will follow to go away from its bifurcation state). 
The dynamical equation for the amplitude $\mathcal{A}(\tau)$ is obtained from a solvability condition for the system at the next order.

\bigskip
\par\noindent
\textbf{At Order $\Delta \mu$.} 
The acceleration term comes into play at this order where we have
\begin{equation}
\begin{split}
\displaystyle \mathcal{L}(W_1, \Lambda_1)=\underbrace{-\Phi_0\frac{d^2 \mathcal{A}}{d\tau^2}-\mathcal{A}^2\left(\frac{d^2 W_{\textrm{eq}}^*}{dX^2}+2\Lambda_{\textrm{eq}}^*\frac{d^2\Phi_0}{dX^2}\right)}_{\displaystyle \mathcal{F}_{1}},
\qquad
&\displaystyle \int_{0}^{1}\frac{d W_{\textrm{eq}}^*}{dX}\frac{\partial W_1}{\partial X}dX=\displaystyle -\frac{1}{2}\mathcal{A}^2\int_{0}^{1}\left(\frac{d\Phi_0}{dX}\right)^2dX,
\\
\left. W_1\right|_{X=0}= \left. W_1\right|_{X=1}=0,
\qquad 
&\displaystyle\left.\frac{\partial W_1}{\partial X}\right|_{X=0}=1,
\qquad
\displaystyle\left.\frac{\partial W_1}{\partial X}\right|_{X=1}=0.
\end{split}
\label{eq:secondOrderSystemsymmetric}
\end{equation}
This system is similar to the one obtained at leading order but with a non-homogeneous terms on the right-hand side and non-homogeneous boundary conditions. This system admits a bounded solution only if the right hand side (in the extended sense) of the PDE in \eqref{eq:secondOrderSystemsymmetric} is orthogonal to the adjoint solution \cite{friedman1990,keener1995}. The operator $\mathcal{L}$ is self-adjoint relative to the standard Cartesian scalar product $\langle u,v \rangle=\int_{0}^{1}uvdx$ and the adjoint solution is therefore $\Phi_0(X)$. The solvability condition is then obtained by requiring that
\begin{equation}
\langle \mathcal{F}_1, \Phi_0\rangle-B(W_1, \Phi_0)=0,
\end{equation}
where $\mathcal{F}_1$ is the right hand side of the PDE in~\eqref{eq:secondOrderSystemsymmetric} governing the system at this order, and $B(W_1, \Phi_0)$ is the boundary term obtained by transferring the operator onto the adjoint solution using integration by part \cite{friedman1990,keener1995}.
For the asymmetric case, we obtain the condition 
\begin{equation}
\frac{d^2 \mathcal{A}}{d\tau^2}=a_1+a_2\mathcal{A}^2,
\label{eq:solvCondAsym}
\end{equation}
where $a_1$ and $a_2$ are two positive constants given by 
\begin{equation}
a_1=\frac{4\Lambda_{\textrm{eq}}^*}{\mu^* I_1}, \qquad a_2=\frac{3\Lambda_{\textrm{eq}}^* I_2}{I_1},
\end{equation}
and $I_1$ and $I_2$ are 
\begin{equation}
I_1=\int_{0}^{1}\Phi_0^2 dx, \qquad 
I_2=\int_{0}^{1}\left(\frac{d\Phi_0}{dx}\right)^2 dx.
\end{equation}
Eqn. \eqref{eq:solvCondAsym} is the one obtained by Gomez \textrm{et al.}\cite{gomez2017} to describe the snapping dynamics observed when the system is placed to the right of the bifurcation ($\mathcal{A}$ and $\tau$ correspond to the relevant $O(1)$ amplitude and time when approaching the bifurcation).  It corresponds to the canonical form of a saddle node bifurcation. It does not display the bifurcation parameter as the latter is actually hidden in the rescaled amplitude $\mathcal{A}$ and time $\tau$.

\end{document}